\documentstyle[12pt]{article}

\font\twlgot =eufm10 scaled \magstep1 \font\egtgot =eufm8
\font\sevgot =eufm7 \font\twlmsb =msbm10 scaled \magstep1
\font\egtmsb =msbm8 \font\sevmsb =msbm7

\newfam\gotfam
\def\pgot{\fam\gotfam\twlgot}
\textfont\gotfam\twlgot \scriptfont\gotfam\egtgot
\scriptscriptfont\gotfam\sevgot
\def\got{\protect\pgot}
\newfam\msbfam
\textfont\msbfam\twlmsb \scriptfont\msbfam\egtmsb
\scriptscriptfont\msbfam\sevmsb
\def\Bbb{\protect\pBbb}
\def\pBbb{\relax\ifmmode\expandafter\Bb\else\typeout{You cann't use
Bbb in text mode}\fi}
\def\Bb #1{{\fam\msbfam\relax#1}}

\newcommand{\gQ}{{\got T}}

\newcommand{\gA}{{\got A}}

\newcommand{\gd}{{\got d}}

\newcommand{\gS}{{\got S}}

\newcommand{\gL}{{\got L}}

\def\thebibliography#1{\section*{References}\list
  {[\arabic{enumi}]}{\settowidth\labelwidth{#1}\leftmargin\labelwidth
    \advance\leftmargin\labelsep
    \usecounter{enumi}}
    \def\newblock{\hskip .11em plus .33em minus .07em}
    \sloppy\clubpenalty4000\widowpenalty4000
    \sfcode`\.=1000\relax}

\def\op#1{\mathop{\fam0 #1}\limits}

\newcommand{\im}{{\rm Im\,}}
\newcommand{\nm}[1]{|{#1}|}
\newcommand{\beq}{\begin{equation}}
\newcommand{\eeq}{\end{equation}}
\newcommand{\ben}{\begin{eqnarray}}
\newcommand{\een}{\end{eqnarray}}
\newcommand{\be}{\begin{eqnarray*}}
\newcommand{\ee}{\end{eqnarray*}}
\newcommand{\bea}{\begin{eqalph}}
\newcommand{\eea}{\end{eqalph}}
\newcommand{\cA}{{\cal A}}

\newcommand{\cP}{{\cal P}}

\newcommand{\cL}{{\cal L}}

\newcommand{\cE}{{\cal E}}

\newcommand{\cS}{{\cal S}}
\newcommand{\cC}{{\cal C}}
\newcommand{\cO}{{\cal O}}

\newcommand{\bL}{{\bf L}}

\newcommand{\bE}{{\bf E}}

\newcommand{\al}{\alpha}
\newcommand{\vr}{\varrho}
\newcommand{\bt}{\beta}
\newcommand{\dl}{\delta}
\newcommand{\la}{\lambda}
\newcommand{\La}{\Lambda}
\newcommand{\f}{\phi}
\newcommand{\om}{\omega}

\newcommand{\m}{\mu}

\newcommand{\g}{\gamma}
\newcommand{\G}{\Gamma}
\newcommand{\th}{\theta}

\newcommand{\vt}{\vartheta}

\newcommand{\up}{\upsilon}

\newcommand{\di}{{\rm dim\,}}

\newcommand{\si}{\sigma}
\newcommand{\Si}{\Sigma}
\newcommand{\w}{\wedge}

\newcommand{\ol}{\overline}
\newcommand{\dr}{\partial}
\newcommand{\ar}{\op\longrightarrow}
\newcommand{\ot}{\otimes}

\newcommand{\e}{\epsilon}

\newcommand{\rdr}{\stackrel{\leftarrow}{\dr}{}}
\newcommand{\lto}{\leftarrow}
\newcommand{\llr}{\op\longleftarrow}

\let\ssection=\section
\renewcommand{\section}{\setcounter{equation}{0}\ssection}

\newcounter{example}[section]
\newcounter{remark}[section]
\newcounter{theorem}[section]
\newcounter{proposition}[section]
\newcounter{lemma}[section]
\newcounter{corollary}[section]
\newcounter{definition}[section]

\def\theremark{\arabic{section}.\arabic{remark}}

\def\thedefinition{\arabic{section}.\arabic{definition}}

\newenvironment{proof}{\noindent
{\it Proof.}}{$\Box$ \medskip}
\newenvironment{rem}{\refstepcounter{remark}\medskip\noindent{\it
Remark \theremark.}}{\medskip}
\newenvironment{ex}{\refstepcounter{remark}\medskip\noindent{\it
Example \theremark.}}{\medskip}
\newenvironment{theo}{\refstepcounter{definition}
\bigskip\noindent{\bf Theorem \thedefinition.} \it}{\medskip}
\newenvironment{prop}{\refstepcounter{definition}
\bigskip\noindent{\bf Proposition \thedefinition.}\it}{\medskip}
\newenvironment{lem}{\refstepcounter{definition}
\bigskip\noindent{\bf Lemma \thedefinition.}\it}{\medskip}
\newenvironment{cor}{\refstepcounter{definition}
\bigskip\noindent{\bf Corollary \thedefinition.}\it}{\medskip}
\newenvironment{defi}{\refstepcounter{definition}
\bigskip\noindent{\bf Definition \thedefinition.}\it}{\medskip}

\newcommand{\mar}[1]{}

\begin{document}
\hbox{}

{\parindent=0pt

{\large \bf On necessary and sufficient conditions of the BV
quantization of a generic Lagrangian field system}
\bigskip

{\bf Denis Bashkirov}$^1$, {\bf Giovanni Giachetta}$^2$, {\bf
Luigi Mangiarotti}$^2$, {\bf Gennadi Sardanashvily}$^1$
\bigskip

\begin{small}

$^1$ Department of Theoretical Physics, Physics Faculty, Moscow
State University, 117234 Moscow, Russia
\medskip

$^2$ Department of Mathematics and Informatics, University of
Camerino, 62032 Camerino (MC), Italy

\bigskip

\end{small}

{\bf Abstract:} We address the problem of extending an original field
Lagrangian to ghosts and antifields in order to satisfy the master
equation in the framework of the BV quantization of Lagrangian field
systems. This extension essentially depends on the degeneracy of an
original Lagrangian whose Euler--Lagrange operator generally obeys the
Noether identities which need not be independent, but satisfy the
first-stage Noether identities, and so on. A generic Lagrangian system of
even and odd fields on an arbitrary smooth manifold  is examined in the
algebraic terms of the Grassmann-graded variational bicomplex. 
We state the necessary and sufficient condition for the existence of the
exact antifield Koszul--Tate complex whose boundary operator
provides all the Noether and higher-stage Noether identities of an
original Lagrangian system. The Noether inverse second theorem that we
prove  associates to this Koszul--Tate complex the sequence of
ghosts whose ascent operator provides the gauge and higher-stage gauge
supersymmetries of an original Lagrangian. We show that an original
Lagrangian is extended to a solution of the master equation if this
ascent operator admits a nilpotent extension and only if it is extended
to an operator nilpotent on the shell.  

  }

\section{Introduction}

As is well known, the Batalin-Vilkovisky (henceforth BV) quantization of a
Lagrangian field system essentially depends on the analysis of its
degeneracy \cite{barn,bat,fust,gom}. A Lagrangian
system is said to be degenerate if its Euler--Lagrange operator obeys
non-trivial Noether identities. They need not be independent, but
satisfy the first-stage Noether identities, which in turn are
subject to the second-stage ones, and so on. The hierarchy of
reducible Noether identities characterizes the degeneracy of a
Lagrangian system in full. The Noether second theorem states the
relation between the Noether identities and the gauge symmetries
of a Lagrangian system. If Noether identities and gauge symmetries
are finitely generated, they are parameterized by the modules of
antifields and ghosts, respectively. An original Lagrangian is
extended to these antifields and ghosts in order to satisfy the
so-called classical master equation. This extended Lagrangian is the main
ingredient in the BV quantization procedure.

It should be noted that the notion of reducible Noether identities
has come from that of reducible constraints. Their Koszul--Tate
complex has been invented by analogy with that of constraints
\cite{fisch} under a rather restrictive regularity condition that
field equations as well as Noether identities of arbitrary stage
can be locally separated into the independent and dependent ones
\cite{barn,fisch}. This condition has also come from the case of
constraints locally given by a finite number of functions which
the inverse mapping theorem is applied to. A problem is that, 
in contrast
with constraints, Noether and higher-stage Noether identities are
differential operators. They are locally given by a set of
functions and their jet prolongations on an infinite order jet
manifold. Since the latter is a Fr\'echet, but not Banach
manifold, the inverse mapping theorem fails to be valid.

Following the general notion of Noether identities of differential
operators \cite{jmp05a,oper}, we here address a generic degenerate
Lagrangian system of even and odd fields on an arbitrary smooth manifold.
Dealing with odd fields, we follow the algebraic description of Lagrangian
systems in terms of the Grassmann-graded variational bicomplex 
\cite{barn,jmp05,cmp04}. 
Theorem \ref{v11} provides its relevant cohomology.

We show that, if Noether and higher-stage Noether identities are
assumed to be finitely generated and iff a certain homology regularity
condition (Definition
\ref{v155}) holds, one can associate to the Euler--Lagrange operator of a
degenerate Grassmann-graded Lagrangian system the exact Koszul--Tate
complex (\ref{w3}) of antifields whose boundary operator (\ref{w4})
provides all the Noether and higher-stage Noether identities of an
original Lagrangian system (Theorem
\ref{v163}).

The Noether second theorem relates the Noether and higher-stage
Noether identities to the gauge and higher-stage gauge symmetries
and supersymmetries of a Lagrangian system
\cite{jpa05,jmp05,fulp}. We prove its variant (Theorem \ref{w35}) which
associates to the
above mentioned Koszul--Tate complex the sequence (\ref{w36}), graded in
ghosts, whose ascent operator (\ref{w108}) provides gauge and
higher-stage gauge supersymmetries of an original 
Lagrangian system. We agree to call it the total gauge operator, acting
both on original fields and ghosts. This operator need not be nilpotent. 
If it admits a nilpotent (resp. nilpotent on the shell) extension, one
can say that gauge and higher-stage gauge supersymmetries of an original
Lagrangian system constitute an algebra (resp. an algebra on the shell). 

Extending an original Lagrangian system to the above mentioned ghosts and
antifields, we come to a Lagrangian system whose Lagrangian
can satisfy the particular condition called the classical master equation
(Proposition \ref{w39}).  We show that an original Lagrangian is extended to a
nontrivial solution of the master equation only if the above mentioned
total gauge operator (\ref{w108})  is extended to an operator
nilpotent on the shell (Theorem \ref{w120}) and if this operator  
admits a
nilpotent extension (Theorem \ref{w130}).

Note that the proof of Theorem \ref{w130} appeals to the above mentioned
homology regularity condition, and states something more. Given a
nilpotent extension (\ref{w109}) of the total gauge operator, a desired
solution of the master equation is obtained at once by the formula
(\ref{w133}) (or (\ref{w300})). It is affine in antifields, and this fact
is essential for the further gauge fixing procedure. We thus may conclude
that  the study of a classical Lagrangian field system for
the purpose of its BV quantization mainly reduces to constructing the
total gauge operator (\ref{w108}) and its nilpotent extension. 

In Section 6, an example coming from  topological BF theory is examined in
detail. It is a reducible degenerate Lagrangian system whose total gauge
operator is nilpotent and, thus, provides an extension of the original
Lagrangian (\ref{v182}) of the topological BF theory to the solution
(\ref{w302}) of the master equation.

\section{Grassmann-graded Lagrangian systems}

We describe Lagrangian systems of even and
odd variables in algebraic terms of the Grassmann-graded variational
bicomplex
\cite{barn,jmp05,cmp04}, generalizing the well-known variational
bicomplex for even Lagrangian systems on fiber bundles
\cite{ander,jmp,tak2}.

\begin{rem}
Smooth manifolds throughout are real, finite-dimensional,
Hausdorff, second-countable (hence, paracompact) and connected.
Graded manifolds with structure sheaves of Grassmann algebras of
finite rank are only considered. By $\La$, $\Si$, $\Xi$, are
denoted symmetric multi-indices, e.g., $\La=(\la_1...\la_k)$,
$\la+\La=(\la\la_1...\la_k)$. By a summation over a multi-index
$\La=(\la_1...\la_k)$ is meant separate summation over each index
$\la_i$.
\end{rem}

Let $Y\to X$, $\di X=n$, be a fiber bundle.  The jet manifolds $J^rY$ of
its sections form the inverse system
\mar{5.10}\beq
X\op\longleftarrow^\pi Y\op\longleftarrow^{\pi^1_0} J^1Y
\longleftarrow \cdots J^{r-1}Y \op\longleftarrow^{\pi^r_{r-1}}
J^rY\longleftarrow\cdots, \label{5.10}
\eeq
where $\pi^r_{r-1}$  are affine bundles. Its projective limit
$(J^\infty Y;\pi^\infty_r:J^\infty Y\to J^rY)$ is a paracompact
Fr\'echet manifold.  A bundle atlas $\{(U_Y;x^\la,y^i)\}$ of $Y\to
X$ induces the coordinate atlas
\mar{jet1}\ben
&& \{((\pi^\infty_0)^{-1}(U_Y); x^\la, y^i_\La)\}, \qquad
{y'}^i_{\la+\La}=\frac{\dr x^\m}{\dr x'^\la}d_\m y'^i_\La, \qquad
0\leq|\La|, \label{jet1} \\
&& d_\la = \dr_\la + \op\sum_{0\leq|\La|} y^i_{\la+\La}\dr_i^\La,
\qquad d_\La=d_{\la_1}\circ\cdots\circ d_{\la_k}, \nonumber
\een
of $J^\infty Y$, where $d_\la$ are total derivatives. The inverse
system (\ref{5.10}) yields the direct system
\mar{5.7}\beq
\cO^*X\op\longrightarrow^{\pi^*} \cO^*Y
\op\longrightarrow^{\pi^1_0{}^*} \cO_1^*Y \ar\cdots \cO^*_{r-1}Y
\op\longrightarrow^{\pi^r_{r-1}{}^*}
 \cO_r^*Y \longrightarrow\cdots  \label{5.7}
\eeq
of graded differential algebras (henceforth GDAs) $\cO_r^*Y$ of
exterior forms on $X$, $Y$ and jet manifolds $J^rY$ with respect to the
pull-back monomorphisms $\pi^r_{r-1}{}^*$. Its direct limit is the
GDA  $\cO_\infty^*Y$ of all exterior forms on finite order jet
manifolds modulo the pull-back identification. The GDA
$\cO_\infty^*Y$ is split into the above mentioned variational bicomplex
describing Lagrangian systems of even fields on a fiber bundle
$Y\to X$.

Treating odd fields, we appeal to forthcoming Theorem \ref{v0}, which
is a corollary of the Batchelor theorem \cite{bart}
and the Serre-Swan theorem for an arbitrary smooth
manifold \cite{book05,ren}.

\begin{theo} \label{v0} \mar{v0}
A Grassmann algebra $\cA$ over the
ring $C^\infty(Z)$ of smooth real functions on a manifold $Z$ is
isomorphic to the algebra of graded functions on a graded manifold with a
body $Z$ iff it is the exterior algebra of some projective
$C^\infty(Z)$-module of finite rank \cite{jmp05a}.
\end{theo}

Recall that the above mentioned Batchelor theorem states an isomorphism of
a  graded manifold $(Z,\gA)$ with a body $Z$ to the particular one
$(Z,\gA_Q)$ with the structure sheaf $\gA_Q$ of germs of sections
of the exterior bundle
\be
\w Q^*=\Bbb R\op\oplus_Z Q^*\op\oplus_Z\op\w^2
Q^*\op\oplus_Z\cdots,
\ee
where $Q^*$ is the dual of some vector bundle $Q\to Z$. In field models,
Batchelor's isomorphism is usually fixed from the beginning. Let us
call $(Z,\gA_Q)$ the simple graded manifold modelled over $Q$.
Its ring $\cA_Q$ of graded functions 
consists of sections of $\w Q^*$. The following bigraded differential
algebra (henceforth BGDA)
$\cS^*[Q;Z]$ is associated to $(Z,\gA_Q)$ \cite{bart,book05}.

Let $\gd\gA_Q$ be the sheaf of graded derivations of $\gA_Q$. Its
global sections make up the real Lie superalgebra $\gd\cA_Q$ of
(left) graded derivations of the $\Bbb R$-ring $\cA_Q$, i.e.,
\be
u(ff')=u(f)f'+(-1)^{[u][f]}fu(f'), \qquad f,f'\in \cA_Q, \qquad
u\in \gd\cA_Q,
\ee
where the symbol $[.]$ stands for the Grassmann parity. Then the
Chevalley--Eilenberg complex of $\gd\cA_Q$ with coefficients in
$\cA_Q$ can be constructed \cite{fuks}.  Its subcomplex
$\cS^*[Q;Z]$ of $\cA_Q$-linear morphisms is the Grassmann-graded
Chevalley--Eilenberg differential calculus
\be
0\to \Bbb R\to \cA_Q \ar^d \cS^1[Q;Z]\ar^d\cdots
\cS^k[Q;Z]\ar^d\cdots
\ee
over a $\Bbb Z_2$-graded commutative $\Bbb R$-ring $\cA_Q$ 
\cite{book05}. The graded exterior product $\w$ and the
Chevalley--Eilenberg coboundary operator $d$ make $\cS^*[Q;Z]$ into a BGDA
\mar{v22}\beq
\f\w\f' =(-1)^{|\f||\f'| +[\f][\f']}\f'\w \f, \qquad  d(\f\w\f')=
d\f\w\f' +(-1)^{|\f|}\f\w d\f',  \label{v22}
\eeq
where $|.|$ denotes the form degree. 
 Note that $\cS^*[Q;Z]$ is a minimal differential
calculus over
$\cA_Q$, i.e., it is generated by elements $df$, $f\in \cA_Q$. There is
the natural monomorphism $\cO^*Z\to
\cS^*[Q;Z]$.

One can think of elements of the BGDA $\cS^*[Q;Z]$ as being graded
exterior forms on a manifold $Z$ as follows. Given an open subset
$U\subset Z$, let $\cA_U$ be the Grassmann algebra of sections of the
sheaf
$\gA_Q$ over $U$, and let $\cS^*[Q;U]$ be the Chevalley--Eilenberg
differential calculus over $\cA_U$. With another open subset $U'\subset
U$, the restriction morphism $\cA_U\to\cA_{U'}$ yields a
homomorphism  of BGDAs $\cS^*[Q;U]\to \cS^*[Q;U']$. Thus,  we
obtain the presheaf $\{U,\cS^*[Q;U]\}$ of BGDAs on a manifold $Z$
and the sheaf $\gS^*[Q;Z]$ of BGDAs of germs of this presheaf.
Since $\{U,\cA_U\}$ is the canonical presheaf of $\gA_Q$, the
canonical presheaf of $\gS^*[Q;Z]$ is $\{U,\cS^*[Q;U]\}$. In
particular, $\cS^*[Q;Z]$ is the BGDA of global sections of the
sheaf $\gS^*[Q;Z]$, and  there is the restriction morphism
$\cS^*[Q;Z]\to \cS^*[Q;U]$ for any open $U\subset Z$.

Due to this restriction
morphism, elements of $\cS^*[Q;Z]$ can be written in the following
local form. Given bundle coordinates $(z^A,q^a)$ on $Q$ and
the corresponding fiber basis $\{c^a\}$ for $Q^*\to X$, the tuple $(z^A,
c^a)$ is called a local basis for the graded manifold $(Z,\gA_Q)$. With
respect to this basis, graded functions read
\mar{v23}\beq
f=\op\sum_{0\leq k} \frac1{k!}f_{a_1\ldots a_k}c^{a_1}\cdots c^{a_k}.
\qquad f_{a_1\ldots a_k}\in C^\infty(Z). \label{v23}
\eeq
 Due to the canonical splitting $VQ= Q\times Q$,
the fiber basis $\{\dr_a\}$ for vertical tangent bundle $VQ\to Q$
of $Q\to Z$ is the dual of $\{c^a\}$. Then graded derivations take
the local form $u= u^A\dr_A + u^a\dr_a$, where $u^A, u^a$ are
local graded functions. They act on graded functions (\ref{v23})
by the rule
\be
u(f_{a\ldots b}c^a\cdots c^b)=u^A\dr_A(f_{a\ldots b})c^a\cdots c^b
+u^d f_{a\ldots b}\dr_d\rfloor (c^a\cdots c^b). 
\ee
Relative to the dual bases $\{dz^A\}$ for $T^*Z$ and
$\{dc^b\}$ for $Q^*$, graded one-forms read $\f=\f_A dz^A +
\f_adc^a$. The duality morphism and the graded exterior differential
$d$ take the form
\be
 u\rfloor \f=u^A\f_A + (-1)^{[\f_a]}u^a\f_a, \qquad 
d\f=dz^A\w \dr_A\f + dc^a\w \dr_a\f.
\ee

\begin{rem} \label{w11} \mar{w11}
One also deals with right graded derivations $\op u^\lto$ of graded
functions and the right graded exterior differential $\op d^\lto$. They
read
\be
&& \op u^\lto(ff')=f\op u^\lto(f')+(-1)^{[\op u^\lto][f']}\op
u^\lto(f)f', \qquad f,f'\in \cA_Q, \\
&& \op u^\lto(f)= \dr_A (f)u^A + {\op\dr^\lto}_d(f) u^d, \qquad
{\op\dr^\lto}_d(f_{a\ldots b}c^a\cdots c^b)= f_{a\ldots b} (c^a\cdots
c^b)\lfloor {\op\dr^\lto}_d, \\
&& \op d^\lto(\f)=\dr_A(\f)\w dz^A + {\op\dr^\lto}_a(\f)\w dc^a,
\qquad \f\in \cS^*[Q;Z].
\ee
\end{rem}

If $Y\to X$ is an affine bundle, the total algebra
of even and odd fields has been defined as the
product of the polynomial subalgebra of the GDA $\cO_\infty^*Y$
and some BGDA of graded exterior forms on graded manifolds whose
body is $X$ \cite{jmp05,cmp04}. Since $Y\to X$ here need not be
affine, we consider graded manifolds whose bodies are 
$Y$ and jet manifolds $J^rY$ \cite{jmp05a}. This definition of jets of
odd fields differs from that of jets of a graded commutative ring
\cite{book05} and jets of a graded fiber bundle \cite{hern}, but
reproduces the heuristic notion of jets of odd ghosts in the
Lagrangian BRST theory \cite{barn,bran01}.

Given a vector bundle $F\to X$, let us consider the simple graded
manifold $(J^rY,\gA_{F_r})$ modelled over the pull-back
$F_r=J^rY\times_XJ^rF$  onto $J^rY$ of the jet bundle $J^rF\to X$.
There is an epimorphism of graded manifolds
\be
(J^{r+1}Y,\gA_{F_{r+1}}) \to (J^rY,\gA_{F_r}),
\ee
regarded as local-ringed spaces. It consists of the surjection
$\pi^{r+1}_r$ and the sheaf monomorphism $\pi_r^{r+1*}\gA_{F_r}\to
\gA_{F_{r+1}}$, where $\pi_r^{r+1*}\gA_{F_r}$ is the pull-back
onto $J^{r+1}Y$ of the topological fiber bundle $\gA_{F_r}\to
J^rY$. This sheaf monomorphism induces the monomorphism of the
canonical presheaves $\ol \gA_{F_r}\to \ol \gA_{F_{r+1}}$, which
associates to each open subset $U\subset J^{r+1}Y$ the ring of
sections of $\gA_{F_r}$ over $\pi^{r+1}_r(U)$. Accordingly, there
is the monomorphism of graded commutative rings $\cA_{F_r} \to
\cA_{F_{r+1}}$. Then this monomorphism yields the monomorphism of BGDAs
\mar{v4}\beq
\cS^*[F_r;J^rY]\to \cS^*[F_{r+1};J^{r+1}Y]. \label{v4}
\eeq
As a consequence, we have the direct system of BGDAs
\mar{v6}\beq
\cS^*[Y\op\times_X F;Y]\ar \cS^*[F_1;J^1Y]\ar\cdots
\cS^*[F_r;J^rY]\ar\cdots, \label{v6}
\eeq
whose direct limit $\cS^*_\infty[F;Y]$  is a BGDA of all graded
differential forms on jet manifolds $J^rY$ modulo monomorphisms
(\ref{v4}). Its elements obey the relations 
(\ref{v22}). The monomorphisms $\cO^*_rY\to \cS^*[F_r;J^rY]$
provide a monomorphism of the direct system (\ref{5.7}) to the
direct system (\ref{v6}) and, consequently, the monomorphism
$\cO^*_\infty Y\to \cS^*_\infty[F;Y]$ of their direct limits. In
particular, $\cS^*_\infty[F;Y]$ is an $\cO^0_\infty Y$-algebra.

It is the BGDA $\cS^*_\infty[F;Y]$ which provides algebraic description
of Lagrangian systems of even and odd fields.
If $Y\to X$ is an affine bundle, we recover the BGDA introduced in
\cite{jmp05,cmp04} by restricting the ring $\cO^0_\infty Y$ to its
subring $\cP^0_\infty Y$ of polynomial functions, but now elements
of $\cS^*_\infty[F;Y]$ are graded exterior forms on
 $J^\infty Y$.
Indeed, let $\gS^*[F_r;J^rY]$ be the sheaf of BGDAs on $J^rY$ and
$\ol\gS^*[F_r;J^rY]$ its canonical presheaf whose elements are the
Chevalley--Eilenberg differential calculus over elements of the
presheaf $\ol\gA_{F_r}$. Then the presheaf monomorphisms $\ol
\gA_{F_r}\to \ol \gA_{F_{r+1}}$ yield the direct system of
presheaves
\mar{v15}\beq
\ol\gS^*[Y\times F;Y]\ar \ol\gS^*[F_1;J^1Y] \ar\cdots
\ol\gS^*[F_r;J^rY]  \ar\cdots, \label{v15}
\eeq
whose direct limit $\ol\gS_\infty^*[F;Y]$ is a presheaf of BGDAs
on the infinite order jet manifold $J^\infty Y$. Let
$\gQ^*_\infty[F;Y]$ be the sheaf of BGDAs of germs of the presheaf
$\ol\gS_\infty^*[F;Y]$.  The structure module
$\G(\gQ^*_\infty[F;Y])$ of sections of $\gQ^*_\infty[F;Y]$ is a
BGDA such that, given an element $\f\in \G(\gQ^*_\infty[F;Y])$ and
a point $z\in J^\infty Y$, there exist an open neighbourhood $U$
of $z$ and a graded exterior form $\f^{(k)}$ on some finite order
jet manifold $J^kY$ so that $\f|_U= \pi^{\infty*}_k\f^{(k)}|_U$.
In particular, there is the  monomorphism $\cS^*_\infty[F;Y]
\to\G(\gQ^*_\infty[F;Y])$.

Due to this monomorphism, one can restrict $\cS^*_\infty[F;Y]$ to
the coordinate chart (\ref{jet1}) and say that $\cS^*_\infty[F;Y]$
as an $\cO^0_\infty Y$-algebra is locally generated by  the
elements
\be
(1, c^a_\La,
dx^\la,\th^a_\La=dc^a_\La-c^a_{\la+\La}dx^\la,\th^i_\La=
dy^i_\La-y^i_{\la+\La}dx^\la), \qquad 0\leq |\La|.
\ee
One call $(y^i,c^a)$ the local basis for $\cS^*_\infty[F;Y]$. We
further use the collective symbol $s^A$ for its elements, together with
the notation $s^A_\La$, $\th^A_\La=ds^A_\La- s^A_{\la+\La}dx^\la$,
and $[A]=[s^A]$.

\begin{rem} Given
local graded functions $f^\La$ and a graded form $\Phi$, there are
useful relations
\mar{0606b-d}\ben
&& \op\sum_{0\leq |\La|\leq k} (-1)^{|\La|}d_\La(f^\La \Phi)=
\op\sum_{0\leq |\La|\leq k} \eta (f)^\La d_\La \Phi, \label{0606b}
\\ && \eta (f)^\La = \op\sum_{0\leq|\Si|\leq
k-|\La|}(-1)^{|\Si+\La|} C^{|\Si|}_{|\Si+\La|} d_\Si f^{\Si+\La},
\qquad C^a_b=\frac{b!}{a!(b-a)!}, \label{0606c}\\
&& (\eta\circ\eta)(f)^\La=f^\La. \label{0606d}
\een
\end{rem}

The BGDA $\cS^*_\infty[F;Y]$ is decomposed into
$\cS^0_\infty[F;Y]$-modules $\cS^{k,r}_\infty[F;Y]$ of $k$-contact
and $r$-horizontal graded forms
\be
&& \f=\op\sum_{0\leq|\La_i|}\f^{\La_1\ldots \La_k}_{A_1\ldots A_k
\m_1\ldots\m_r}
\th^{A_1}_{\La_1}\w\cdots\w\th^{A_k}_{\La_k}\w dx^{\m_1}\w\cdots\w
dx^{\m_r},
\qquad
\f\in
\cS^{k,r}_\infty[F;Y],\\
&& h_k: \cS^*_\infty[F;Y] \to \cS^{k,*}_\infty[F;Y], \qquad 
h^r: \cS^*_\infty[F;Y] \to \cS^{*,r}_\infty[F;Y].
\ee
Accordingly, the graded exterior differential $d$ on
$\cS^*_\infty[F;Y]$ falls into the sum $d=d_H+d_V$ of the total
and vertical differentials where
\be
d_H\circ h_k=h_k\circ d\circ h_k, 
 \qquad d_H(\f)= dx^\la\w d_\la(\f),  \qquad d_\la = \dr_\la +
\op\sum_{0\leq|\La|} s^A_{\la+\La}\dr_A^\La.
\ee
With the graded projection endomorphism
\be
\vr=\op\sum_{k>0} \frac1k\ol\vr\circ h_k\circ h^n, \qquad
\ol\vr(\f)= \op\sum_{0\leq|\La|} (-1)^{\nm\La}\th^A\w
[d_\La(\dr^\La_A\rfloor\f)], \qquad \f\in \cS^{>0,n}_\infty[F;Y],
\ee
and the graded variational operator $\dl=\vr\circ d$, the BGDA
$\cS^*_\infty[F;Y]$ is split into the above mentioned
Grassmann-graded variational bicomplex. We restrict our
consideration to its short variational subcomplex
and the subcomplex of one-contact graded forms
\mar{g111,2}\ben
&& 0\to \Bbb R\ar \cS^0_\infty[F;Y]\ar^{d_H}\cS^{0,1}_\infty[F;Y]
\cdots \ar^{d_H} \cS^{0,n}_\infty[F;Y]\ar^\dl \bE_1
=\vr(\cS^{1,n}_\infty[F;Y]), \label{g111}\\
&& 0\to \cS^{1,0}_\infty[F;Y]\ar^{d_H} \cS^{1,1}_\infty[F;Y]\cdots
\ar^{d_H}\cS^{1,n}_\infty[F;Y]\ar^\vr \bE_1\to 0. \label{g112}
\een
One can think of their even elements
\mar{0709,'}\ben
&& L=\cL\om\in \cS^{0,n}_\infty[F;Y], \qquad \om=dx^1\w\cdots \w
dx^n,
\label{0709}\\
&& \dl L= \th^A\w \cE_A\om=\op\sum_{0\leq|\La|}
 (-1)^{|\La|}\th^A\w d_\La (\dr^\La_A L)\om\in \bE_1 \label{0709'}
\een
as being a graded Lagrangian and its Euler--Lagrange operator,
respectively.

\begin{rem}
Any graded density $L$ (\ref{0709}) obeys the identity
\be
&& 0=(\dl\circ\dl)(L)=\op\sum_{0\leq|\La|} (-1)^{|\La|}\th^A\w
d_\La(\dr^\La_A\rfloor d(\dl
L))\w\om= \\
&& \qquad \op\sum_{0\leq|\La|}[-\th^A\w\th^B_\La\dr^\La_B\cE_A
+(-1)^{|\La|+[A][B]}\th^B\w d_\La(\th^A\dr^\La_B\cE_A)]\w\om=0,
\ee
which leads to useful equalities
\mar{w51}\beq
 \eta(\dr_B\cE_A)^\La=(-1)^{[A][B]}\dr_A^\La\cE_B \label{w51}
\eeq
(see the notation (\ref{0606c})). It should be noted that 
\be
\eta(\dr_B\cE_A)^{\La=0}=(-1)^{[A][B]}\dr_A\cE_B +S_{AB}, \qquad
S_{AB}=(-1)^{[A][B]}S_{BA},
\ee
but $S_{AB}=0$ because the relation (\ref{0606d}) results in
\be
&&\eta(\eta(\dr_B\cE_A))^{\La=0}=\eta((-1)^{[A][B]}\dr_A\cE_B+S_{AB})^{\La=0}=\\
&& \qquad
(-1)^{[A][B]}\eta(\dr_A\cE_B)^{\La=0}+\eta(S_{AB})^{\La=0}=
\dr_B\cE_A +(-1)^{[A][B]}S_{AB}+ S_{AB}=\dr_B\cE_A.
\ee
\end{rem}

If $Y\to X$ is an affine bundle, cohomology of the complex
(\ref{g111}) equals the de Rham cohomology of $X$, while the
complex (\ref{g112}) is exact \cite{cmp04}. Forthcoming Theorem
\ref{v11} (see Appendix A for its proof) generalizes this result to 
an arbitrary fiber bundle $Y\to X$.  

\begin{theo} \label{v11} \mar{v11}
(i) Cohomology of the complex (\ref{g111}) equals the de Rham
cohomology $H^*(Y)$ of $Y$. (ii) The complex (\ref{g112}) is exact.
\end{theo}

\begin{cor} \label{cmp26} \mar{cmp26}
A $\dl$-closed (i.e., variationally trivial) graded density
$L\in \cS^{0,n}_\infty[F;Y]$ reads
\mar{g215}\beq
L=h_0\psi + d_H\xi, \qquad \xi\in \cS^{0,n-1}_\infty[F;Y],
\label{g215}
\eeq
where $\psi$ is a closed $n$-form on $Y$. In particular, a
$\dl$-closed odd graded density is $d_H$-exact.
\end{cor}

\begin{cor} \label{cmp26'} \mar{cmp26'}
Any graded density $L$ admits the decomposition
\mar{g99}\beq
dL=\dl L - d_H\Xi,
\qquad \Xi\in \cS^{1,n-1}_\infty[F;Y], \label{g99}\\
\eeq
where $L+\Xi$ is a Lepagean equivalent of $L$ \cite{cmp04}.
\end{cor}

The decomposition (\ref{g99})  leads to the first variational
formula (\ref{g107}) for graded Lagrangians \cite{jmp05,cmp04}.
Let $\vt\in\gd \cS^0_\infty[F;Y]$ be a graded derivation of the
$\Bbb R$-ring $\cS^0_\infty[F;Y]$. The interior product
$\vt\rfloor\f$ and the Lie derivative $\bL_\vt\f$,
$\f\in\cS^*_\infty[F;Y]$, are defined by the formulas
\be
&& \vt\rfloor \f=\vt^\la\f_\la + (-1)^{[\f_A]}\vt^A\f_A, \qquad
\f\in \cS^1_\infty[F;Y],\\
&& \vt\rfloor(\f\w\si)=(\vt\rfloor \f)\w\si
+(-1)^{|\f|+[\f][\vt]}\f\w(\vt\rfloor\si), \qquad \f,\si\in
\cS^*_\infty[F;Y], \\
&& \bL_\vt\f=\vt\rfloor d\f+ d(\vt\rfloor\f), \qquad
\bL_\vt(\f\w\si)=\bL_\vt(\f)\w\si
+(-1)^{[\vt][\f]}\f\w\bL_\vt(\si).
\ee
A graded derivation $\vt$ is called contact if the Lie
derivative $\bL_\vt$ preserves the ideal of contact graded forms
of the BGDA $\cS^*_\infty[F;Y]$. Further, we restrict our consideration 
to vertical contact graded derivations, vanishing on $\cO^*X\subset 
\cS^*_\infty[F;Y]$. With respect to the local basis
$\{s^A\}$ for the BGDA $\cS^*_\infty[F;Y]$, any vertical contact graded
derivation takes the form
\mar{0672}\beq
\vt=\up^A\dr_A + \op\sum_{0<|\La|} d_\La\up^A\dr_A^\La,
\label{0672}
\eeq
where the tuple of graded derivations $\{\dr^\La_A\}$ is
defined as the dual of the tuple of contact forms $\{\th^A_\La\}$,
and  $\up^A$ \cite{cmp04}.
Such a derivation is completely determined by its first summand
$\up=\up^A\dr_A$, called a generalized graded vector field. It
satisfies the relations
\be
\vt\rfloor d_H\f=-d_H(\vt\rfloor\f), \qquad
\bL_\vt(d_H\f)=d_H(\bL_\vt\f), \qquad \f\in\cS^*_\infty[F;Y].
\ee
Then it follows from the splitting (\ref{g99}) that the Lie
derivative $\bL_\vt L$ of a Lagrangian $L$ along a vertical
contact graded derivation $\vt$ (\ref{0672}) fulfills the first
variational formula
\mar{g107}\beq
\bL_\vt L= \up\rfloor\dl L +d_H(\vt\rfloor \Xi)). \label{g107}
\eeq
One says that an odd vertical contact graded derivation $\vt$
(\ref{0672}) is a variational supersymmetry of a graded Lagrangian $L$ if
the Lie derivative $\bL_\vt L$ is $d_H$-exact, i.e.,
the odd graded density $\up\rfloor\dl L=\up^A\cE_A\om$ is $d_H$-exact.

\begin{rem} Given local graded functions $f^\La$, $0\leq
|\La|\leq k$, and $f'$, there is 
the useful relation
\mar{0606a}\beq
\op\sum_{0\leq |\La|\leq k} f^\La d_\La f'\om= \op\sum_{0\leq
|\La|\leq k} (-1)^{|\La|}d_\La (f^\La) f'\om + d_H\si.
\label{0606a}
\eeq
The first variational formula (\ref{g107}) takes the local form
(\ref{0606a}), but it follows from Corollary \ref{cmp26}) that the
second term in its right-hand side is globally $d_H$-exact.
\end{rem}

A vertical contact graded derivation $\vt$ (\ref{0672}) is called
nilpotent if $\bL_\vt(\bL_\vt\f)=0$ for any horizontal graded form
$\f\in \cS^{0,*}_\infty[F;Y]$. One can show that $\vt$
(\ref{0672}) is nilpotent only if it is odd and iff all $\up^A$
obey the equality
\mar{0688}\beq
\vt(\up)=\vt(\up^A\dr_A)=\op\sum_{0\leq|\Si|} 
\up^B_\Si\dr^\Si_B(\up^A)\dr_A=0.
\label{0688}
\eeq

For the sake of simplicity, the common symbol further stands for a
generalized vector field $\up$, the contact graded derivation
$\vt$ (\ref{0672}) determined by $\up$  and the Lie derivative
$\bL_\vt$. We agree to call all these operators the graded
derivation of the BGDA $\cS^*_\infty[F;Y]$.

\begin{rem} \label{w74} \mar{w74}
We further deal with right contact graded derivations $\op\up^\lto
={\op\dr^\lto}_A\up^A$ of the BGDA $\cS^*_\infty[F;Y]$ (see Remark
\ref{w11}). They act on graded forms $\f$
on the right by the rule
\be
\op\up^\lto(\f)=\op d^\lto(\f)\lfloor \op\up^\lto +\op
d^\lto(\f\lfloor\op\up^\lto), \qquad
\op\up^\lto(\f\w\f')=(-1)^{[\f'][\op\up^\lto]}\op\up^\lto(\f)\w\f'+
\f\w\op\up^\lto(\f').
\ee
For instance, ${\op\dr^\lto}_A(\f)=(-1)^{([\f]+1)[A]}\dr_A(\f)$, 
${\op d^\lto}_\La=d_\La$ and ${\op d^\lto}_H(\f)=
(-1)^{|\f|}d_H(\f)$.
With right graded derivations, one obtains the right
Euler--Lagrange operator
\be
\op\dl^\lto L= {\op\cE^\lto}_A\om\w \th^A, \qquad {\op\cE^\lto}_A
=\op\sum_{0\leq|\La|}
 (-1)^{|\La|}d_\La (\rdr^\La_A (L)).  
\ee
An odd right graded derivation $\op\up^\lto$
is a variational supersymmetry of a graded Lagrangian $L$ iff the odd
graded  density 
${\op\cE^\lto}_A\up^A\om$ is $d_H$-exact.
\end{rem}

\section{The Koszul--Tate complex}

Given a graded Lagrangian $L$ (\ref{0709}), let us associate
to its
Euler--Lagrange operator $\dl L$ (\ref{0709'}) the exact chain 
Koszul--Tate complex with the boundary
operator whose nilpotency conditions provide the Noether and
higher-stage Noether identities for $\dl L$ \cite{jmp05a}. 
We follow the
general construction of the Koszul--Tate complex for differential
operators \cite{oper}.

\begin{rem}
If there is no danger of confusion, elements of homology are
identified to its representatives. A chain complex is called
$r$-exact if its homology of degree $k\leq r$ is trivial.
\end{rem}

We start with the following notation. Given a vector bundle
$E\to X$ and its pull-back $E_Y$ onto $Y$, let us consider the
BGDA $\cS^*_\infty[F;E_Y]$. There are monomorphisms of
$\cO^0_\infty Y$-algebras
\be
\cS^*_\infty[F;Y]\to \cS^*_\infty[F;E_Y], \qquad \cO^*_\infty E\to
\cS^*_\infty[F;E_Y],
\ee
whose images contain the common subalgebra $\cO^*_\infty Y$. Let
us consider: (i) the subring $\cP^0_\infty E_Y\subset \cO^0_\infty
E_Y$ of polynomial functions in fiber coordinates  of the vector
bundles $J^rE_Y\to J^rY$, (ii) the corresponding
subring $\cP^0_\infty[F;E_Y]\subset\cS^0_\infty[F;E_Y]$ of graded
functions with polynomial coefficients belonging to $\cP^0_\infty
E_Y$, (iii) the  subalgebra  $\cP^*_\infty[F;Y;E]$ of the BGDA
$\cS^*_\infty[F;E_Y]$ over the subring $\cP^0_\infty[F;E_Y]$.
Given vector bundles $V,V',E,E'$ over $X$, we further use the
notation
\mar{v90}\beq
\cP^*_\infty[V'V;F;Y;EE']= \cP^*_\infty[V'\op\times_X
V\op\times_XF;Y;E\op\times_X E']. \label{v90}
\eeq
By a density-dual of a vector bundle $E\to X$ is meant $\ol
E^*=E^*\ot_X\op\w^n T^*X$.

\begin{prop} \label{w46} \mar{w46}
The BGDA $\cP^*_\infty[F;Y;E]$ and, similarly, the BGDA
(\ref{v90}) possess the same  cohomology as $\cS^*_\infty[F;Y]$ in
Theorem (\ref{v11}).
\end{prop}

\begin{proof} Since $H^*(Y)=H^*(E_Y)$, this cohomology of the BGDA
$\cS^*_\infty[F;Y]$ equals that of the BGDA $\cS^*_\infty[F;E_Y]$.
Furthermore, one can replace the BGDA $\cS^*_\infty[F;E_Y]$ with
$\cP^*_\infty[F;Y;E]$ in the condition of Theorem (\ref{v11}) due
to the fact that sheaves of $\cP^0_\infty E_Y$-modules are also
sheaves of $\cO^0_\infty Y$-modules.
\end{proof}

\begin{rem} \label{w64} \mar{w64}
In a general setting, one must require that transition functions
of fiber bundles over $Y$ under consideration do not vanish on the
shell. For the sake of simplicity, we here restrict our
consideration to fiber bundles over $Y$ which are the pull-back
onto $Y$ of fiber bundles over $X$. In particular, 
a fiber bundle $Y\to X$ of even fields is assumed to
admit the vertical splitting $VY=Y\times_X W$, where $W$ is a
vector bundle over $X$. Let $\ol Y^*$ denote the density-dual of
$W$ in this splitting.
\end{rem}

\begin{prop} \label{v120} \mar{v120} One can associate to a graded
Lagrangian $L$ (\ref{0709}) the chain complex (\ref{v42}) whose
boundaries vanish on the shell.
\end{prop}

\begin{proof}
Let us extend the BGDA $\cS^*_\infty[F;Y]$ to the BGDA
$\cP^*_\infty[\ol Y^*;F;Y;\ol F^*]$ whose local basis is $\{s^A,
\ol s_A\}$, where  $[\ol s_A]=([A]+1){\rm mod}\,2$. We call $\ol s_A$ the
antifields of antifield number Ant$[\ol s_A]= 1$. The BGDA
$\cP^*_\infty[\ol Y^*;F;Y;\ol F^*]$ is provided with the nilpotent right
graded derivation $\ol\dl=\rdr^A \cE_A$, 
where $\cE_A$ are the graded variational derivatives (\ref{0709'}).
We call $\ol\dl$ the Koszul--Tate differential, and say
that an element $\f\in \cP^*_\infty[\ol Y^*;F;Y;\ol F^*]$ vanishes
on the shell if it is $\ol\dl$-exact, i.e., $\f=\ol\dl\si$. Let us
consider the module $\cP^{0,n}_\infty[\ol Y^*;F;Y;\ol F^*]$ of
graded densities. It is split into the chain complex
\mar{v42}\beq
0\lto \cS^{0,n}_\infty[F;Y] \llr^{\ol\dl} \cP^{0,n}_\infty[\ol
Y^*;F;Y;\ol F^*]_1\cdots \llr^{\ol\dl} \cP^{0,n}_\infty[\ol
Y^*;F;Y;\ol F^*]_k \cdots \label{v42}
\eeq
graded by the antifield number. Its boundaries, by definition,
vanish on the shell.
\end{proof}

Since the
homology $H_{k\neq 1}(\ol\dl)$ 
is not essential for our consideration, let us replace the
complex (\ref{v42}) with the finite one
\mar{v042}\beq
0\lto \im\ol\dl \llr^{\ol\dl} \cP^{0,n}_\infty[\ol Y^*;F;Y;\ol
F^*]_1 \llr^{\ol\dl} \cP^{0,n}_\infty[\ol Y^*;F;Y;\ol F^*]_2.
\label{v042}
\eeq
It is exact at
$\im\ol\dl$, and  its first homology coincides with that of  the
complex (\ref{v42}). Let us consider this homology.
A generic one-chain of the complex (\ref{v042}) takes the form
\mar{0712}\beq
\Phi= \op\sum_{0\leq|\La|} \Phi^{A,\La}\ol s_{\La A} \om, \qquad
\Phi^{A,\La}\in \cS^0_\infty[F;Y], \label{0712}
\eeq
and the cycle condition $\ol\dl \Phi=0$ reads
\mar{0713}\beq
\op\sum_{0\leq|\La|} \Phi^{A,\La} d_\La \cE_A \om=0. \label{0713}
\eeq
 One can think of this equality as being a reduction condition on 
the graded variational derivatives $\cE_A$. Conversely, any
reduction condition of form (\ref{0713}) comes from some cycle
(\ref{0712}). The reduction condition (\ref{0713}) is trivial  if
a cycle is a boundary, i.e., it takes the form
\mar{v44}\beq
\Phi= \op\sum_{0\leq|\La|,|\Si|} T^{(A\La)(B\Si)}d_\Si\cE_B\ol
s_{\La A}\om, \qquad T^{(A\La)(B\Si)}=-(-1)^{[A][B]}
T^{(B\Si)(A\La)}. \label{v44}
\eeq
A Lagrangian system is called degenerate if there
exist non-trivial reduction conditions (\ref{0713}), called the
Noether identities.

One can say something more if the $\cS^0_\infty[F;Y]$-module
$H_1(\ol \dl)$ is finitely generated, i.e., it possesses the
following particular structure. There are elements $\Delta\in
H_1(\ol \dl)$ making up a $\Bbb Z_2$-graded projective
$C^\infty(X)$-module $\cC_{(0)}$ of finite rank which, by virtue
of Theorem \ref{v0}, is isomorphic to the module of
sections of the product $\ol V^*\op\times_X \ol E^*$ of the
density-duals of some vector bundles $V\to X$ and $E\to X$. Let
$\{\Delta_r\}$ be local bases for this $C^\infty(X)$-module. Every
element $\Phi\in H_1(\ol \dl)$ factorizes
\mar{v63,71}\ben
&& \Phi= \op\sum_{0\leq|\Xi|} G^{r,\Xi} d_\Xi \Delta_r\om, \qquad
G^{r,\Xi}\in
\cS^0_\infty[F;Y], \label{v63}\\
&&\Delta_r=\op\sum_{0\leq|\La|} \Delta_r^{A,\La}\ol s_{\La A},
\qquad \Delta_r^{A,\La}\in \cS^0_\infty[F;Y], \label{v71}
\een
via elements of $\cC_{(0)}$, i.e., any Noether identity
(\ref{0713}) is a corollary of Noether identities
\mar{v64}\beq
 \op\sum_{0\leq|\La|} \Delta_r^{A,\La} d_\La \cE_A=0.
\label{v64}
\eeq
Clearly, the factorization (\ref{v63}) is independent of
specification of local bases $\{\Delta_r\}$. We say that the
Noether identities  (\ref{v64}) are complete, and call $\Delta_r\in
\cC_{(0)}$ the Noether operators. 

\begin{ex} \label{w0} \mar{w0}
Let $L$ (\ref{0709}) be a variationally trivial Lagrangian. Its
Euler--Lagrange operator $\dl L=0$ obeys the Noether identities
which are finitely generated by the Noether operators
$\Delta_A=\ol s_A$. For instance, this is the case of the topological
Yang--Mills theory.
\end{ex}

\begin{prop} \label{v137} \mar{v137}
If the homology $H_1(\ol\dl)$ of the complex (\ref{v042}) is
finitely generated, this complex  can be extended to the one-exact
complex (\ref{v66}) with a boundary operator whose nilpotency
conditions are equivalent to the complete Noether identities (\ref{v64}).
\end{prop}

\begin{proof}
Let us enlarge the BGDA $\cP^*_\infty[\ol Y^*;F;Y;\ol F^*]$ to  the
BGDA
\mar{w1}\beq
\cP^*_\infty[\ol E^*\ol Y^*;F;Y;\ol F^*\ol V^*], \label{w1}
\eeq
possessing the local basis $\{s^A,\ol s_A, \ol c_r\}$ where
$[\ol
c_r]=([\Delta_r]+1){\rm mod}\,2$ and Ant$[\ol
c_r]=2$. The BGDA (\ref{w1}) is
provided with the nilpotent right graded derivation
$\dl_0=\ol\dl + \rdr^r\Delta_r$, 
called the zero-stage Koszul--Tate differential. Its nilpotency
conditions (\ref{0688}) are equivalent to the complete Noether
identities (\ref{v64}). Then the module $\cP^{0,n}_\infty[\ol
E^*\ol Y^*;F;Y;\ol F^*\ol V^*]_{\leq 3}$ of graded densities of
antifield number Ant$[\f]\leq 3$ is split into the chain complex
\mar{v66}\ben
&&0\lto \im\ol\dl \llr^{\ol\dl} \cP^{0,n}_\infty[\ol Y^*;F;Y;\ol
F^*]_1\llr^{\dl_0}
\cP^{0,n}_\infty[\ol E^*\ol Y^*;F;Y;\ol F^*\ol V^*]_2 \label{v66}\\
&& \qquad \llr^{\dl_0} \cP^{0,n}_\infty[\ol E^*\ol Y^*;F;Y;\ol
F^*\ol V^*]_3. \nonumber
\een
Let $H_*(\dl_0)$ denote its homology. We have
$H_0(\dl_0)=H_0(\ol\dl)=0$. Furthermore, any one-cycle $\Phi$ up
to a boundary takes the form (\ref{v63}) and, therefore, it is a
$\dl_0$-boundary
\be
\Phi= \op\sum_{0\leq|\Si|} G^{r,\Xi} d_\Xi \Delta_r\om
=\dl_0(\op\sum_{0\leq|\Si|} G^{r,\Xi}\ol c_{\Xi r}\om).
\ee
Hence, $H_1(\dl_0)=0$, i.e., the complex (\ref{v66}) is one-exact.
\end{proof}

Turn now to the homology $H_2(\dl_0)$ of the complex (\ref{v66}).
A generic two-chain  reads
\mar{v77}\beq
\Phi= G + H= \op\sum_{0\leq|\La|} G^{r,\La}\ol c_{\La r}\om +
\op\sum_{0\leq|\La|,|\Si|} H^{(A,\La)(B,\Si)}\ol s_{\La A}\ol
s_{\Si B}\om.
\label{v77}
\eeq
The cycle condition $\dl_0 \Phi=0$ takes the form
\mar{v79}\beq
 \op\sum_{0\leq|\La|} G^{r,\La}d_\La\Delta_r\om +\ol\dl H=0.
\label{v79}
\eeq
One can think of this equality as being the reduction condition
on the Noether operators (\ref{v71}). Conversely, let
\be
\Phi=\op\sum_{0\leq|\La|} G^{r,\La}\ol c_{\La r}\om\in
\cP^{0,n}_\infty[\ol E^*\ol Y^*;F;Y;\ol F^*\ol V^*]_2
\ee
be a graded density such that the reduction condition (\ref{v79})
holds. Obviously, this reduction condition is a cycle condition of
the two-chain (\ref{v77}). The reduction condition (\ref{v79}) is
trivial either if a two-cycle $\Phi$ (\ref{v77}) is a boundary or
its summand $G$, linear in antifields, vanishes on the shell.

A degenerate Lagrangian system in Proposition
\ref{v137} is said to be one-stage reducible if there exist
non-trivial reduction conditions (\ref{v79}), called the
first-stage Noether identities.

\begin{prop} \label{v134} \mar{v134}
First-stage Noether identities can be identified to non-trivial
elements of the homology $H_2(\dl_0)$ iff any $\ol\dl$-cycle
$\f\in \cP^{0,n}_\infty[\ol Y^*;F;Y;\ol F^*]_2$ is a
$\dl_0$-boundary.
\end{prop}

\begin{proof}
It suffices to show that, if the summand $G$ of a two-cycle $\Phi$
(\ref{v77}) is $\ol\dl$-exact, then $\Phi$ is a boundary. If
$G=\ol\dl \Psi$, then
\mar{v169}\beq
\Phi=\dl_0\Psi +(\ol \dl-\dl_0)\Psi + H. \label{v169}
\eeq
The cycle condition reads
\be
\dl_0\Phi=\ol\dl((\ol\dl-\dl_0)\Psi + H)=0.
\ee
Then $(\ol \dl-\dl_0)\Psi + H$ is $\dl_0$-exact since any
$\ol\dl$-cycle $\f\in \cP^{0,n}_\infty[\ol Y^*;F;Y;\ol F^*]_2$, by
assumption, is a $\dl_0$-boundary. Consequently, $\Phi$
(\ref{v169}) is $\dl_0$-exact. Conversely, let $\Phi\in
\cP^{0,n}_\infty[\ol Y^*;F;Y;\ol F^*]_2$ be an arbitrary
$\ol\dl$-cycle.
 The cycle condition reads
\mar{v100}\beq
\ol\dl\Phi= 2\Phi^{(A,\La)(B,\Sigma)}\ol s_{\La A} \ol\dl\ol
s_{\Sigma B}\om= 2\Phi^{(A,\La)(B,\Sigma)}\ol s_{\La A} d_\Si
\cE_B\om=0. \label{v100}
\eeq
It follows that $\Phi^{(A,\La)(B,\Sigma)} \ol\dl\ol s_{\Sigma
B}=0$ for all indices $(A,\La)$. We obtain
\be
\Phi^{(A,\La)(B,\Sigma)} \ol s_{\Sigma B}= G^{(A,\La)(r,\Xi)}d_\Xi
\Delta_r +\ol\dl S^{(A,\La)}.
\ee
Hence, $\Phi$ takes the form
\mar{v135}\beq
\Phi=G'^{(A,\La)(r,\Xi)} d_\Xi\Delta_r \ol s_{\La A}\om +\ol\dl
S^{(A,\La)}\ol s_{\La A}\om. \label{v135}
\eeq
We can associate to $\Phi$ (\ref{v135}) the three-chain
\be
\Psi= G'^{(A,\La)(r,\Xi)} \ol c_{\Xi r} \ol s_{\La A}\om +
S^{(A,\La)}\ol s_{\La A}\om
\ee
such that
\be
\dl_0\Psi=\Phi +\si = \Phi + G''^{(A,\La)(r,\Xi)}d_\La\cE_A \ol
c_{\Xi r} \om + S'^{(A,\La)}\ol\dl\ol s_{\La A}\om.
\ee
Owing to the equality $\ol\dl\Phi=0$, we have $\dl_0\si=0$. Since
the term $G''$ of $\si$ is $\ol\dl$-exact, then $\si$  by
assumption is $\dl_0$-exact, i.e., $\si=\dl_0\psi$. It follow that
$\Phi=\dl_0\Psi -\dl_0\psi$.
\end{proof}

If the condition of Proposition \ref{v134} (called the
two-homology regularity condition) is satisfied, let us assume
that the first-stage Noether identities are finitely generated as
follows. There are elements $\Delta_{(1)}\in H_2(\dl_0)$ making up
a $\Bbb Z_2$-graded projective $C^\infty(X)$-module $\cC_{(1)}$ of
finite rank which is isomorphic to the module of sections of the
product $\ol V^*_1\op\times_X \ol E^*_1$  of the density-duals of
some vector bundles $V_1\to X$ and $E_1\to X$. Let
$\{\Delta_{r_1}\}$ be local bases for this $C^\infty(X)$-module.
Every element $\Phi\in H_2(\dl_0)$ factorizes
\mar{v80,1}\ben
&& \Phi= \op\sum_{0\leq|\Xi|} \Phi^{r_1,\Xi} d_\Xi
\Delta_{r_1}\om, \qquad \Phi^{r_1,\Xi}\in
\cS^0_\infty[F;Y], \label{v80}\\
&&\Delta_{r_1}=G_{r_1}+ h_{r_1}=\op\sum_{0\leq|\La|}
\Delta_{r_1}^{r,\La}\ol c_{\La r} + h_{r_1}, \qquad
 h_{r_1}\om\in
\cP^{0,n}_\infty[\ol Y^*;F;Y;\ol F^*], \label{v81}
\een
via elements of $\cC_{(1)}$, i.e., any first-stage Noether
identity (\ref{v79}) results from the equalities
\mar{v82}\beq
 \op\sum_{0\leq|\La|} \Delta_{r_1}^{r,\La} d_\La \Delta_r +\ol\dl
h_{r_1} =0, \label{v82}
\eeq
called the complete first-stage Noether identities. Elements of
$\cC_{(1)}$ are called the first-stage Noether operators. Note
that first summands $G_{r_1}$ of  operators $\Delta_{r_1}$
(\ref{v81}) are not $\ol\dl$-exact.

\begin{prop} \label{v139} \mar{v139} Given a reducible degenerate
Lagrangian system, let the associated one-exact
complex (\ref{v66}) obey the two-homology regularity condition, and
let its homology $H_2(\dl_0)$ be finitely generated. Then this
complex is extended to the two-exact complex (\ref{v87}) with a
boundary operator whose nilpotency conditions are equivalent to
complete Noether and first-stage Noether identities.
\end{prop}

\begin{proof}
Let us consider the BGDA
$\cP^*_\infty[\ol E^*_1\ol E^*\ol Y^*;F;Y;\ol F^*\ol V^*\ol
V^*_1]$
possessing the local basis $\{s^A,\ol s_A, \ol c_r, \ol c_{r_1}\}$,
where $[\ol c_{r_1}]=([\Delta_{r_1}]+1){\rm mod}\,2$ and Ant$[\ol
c_{r_1}]=3$. It can be provided with the nilpotent graded derivation
$\dl_1=\dl_0 + \rdr^{r_1} \Delta_{r_1}$, called the first-stage 
Koszul--Tate
differential. 
Its nilpotency conditions (\ref{0688}) are equivalent to the complete
Noether identities (\ref{v64}) and complete first-stage Noether
identities (\ref{v82}). Then the module $\cP^{0,n}_\infty[\ol
E^*_1\ol E^*\ol Y^*;F;Y;\ol F^*\ol V^*\ol V^*_1]_{\leq 4}$ of
graded densities of antifield number Ant$[\f]\leq 4$ is split into
the chain complex
\mar{v87}\ben
&&0\lto \im\ol\dl \llr^{\ol\dl} \cP^{0,n}_\infty[\ol Y^*;F;Y;\ol
F^*]_1\llr^{\dl_0} \cP^{0,n}_\infty[\ol E^*\ol Y^*;F;Y;\ol F^*\ol
V^*]_2\llr^{\dl_1}
\label{v87}\\
&& \qquad \cP^{0,n}_\infty[\ol E^*_1\ol E^*\ol Y^*;F;Y;\ol F^*\ol
V^*\ol V^*_1]_3
 \llr^{\dl_1}
\cP^{0,n}_\infty[\ol E^*_1\ol E^*\ol Y^*;F;Y;\ol F^*\ol V^*\ol
V^*_1]_4. \nonumber
\een
Let $H_*(\dl_1)$ denote its homology. It is readily observed that
\be
H_0(\dl_1)=H_0(\ol\dl), \qquad H_1(\dl_1)=H_1(\dl_0)=0.
\ee
By virtue of the expression (\ref{v80}), any two-cycle of the
complex (\ref{v87}) is a boundary
\be
 \Phi= \op\sum_{0\leq|\Xi|} \Phi^{r_1,\Xi} d_\Xi \Delta_{r_1}\om
=\dl_1(\op\sum_{0\leq|\Xi|} \Phi^{r_1,\Xi} \ol c_{\Xi r_1})\om.
\ee
It follows that $H_2(\dl_1)=0$, i.e., the complex (\ref{v87}) is
two-exact.
\end{proof}

If the third homology $H_3(\dl_1)$ of the complex (\ref{v87}) is
not trivial, there are reduction conditions on the first-stage
Noether operators, and so on. 

Iterating the arguments, we
come to the following.
Let $(\cS^*_\infty[F;Y],L)$ be a degenerate 
Lagrangian system whose Noether identities are finitely generated.
In accordance with Proposition \ref{v137}, we associates to it the
one-exact chain complex (\ref{v66}). Given an integer $N\geq 1$,
let $V_1,\ldots, V_N, E_1, \ldots, E_N$ be some vector bundles
over $X$ and 
\mar{v91}\beq
\ol\cP^*_\infty\{N\}=\cP^*_\infty[\ol E^*_N\cdots\ol E^*_1\ol
E^*\ol Y^*;F;Y;\ol F^*\ol V^*\ol V^*_1\cdots\ol V_N^*] \label{v91}
\eeq
the BGDA with the local basis $\{s^A,\ol s_A, \ol c_r, \ol c_{r_1},
\ldots, \ol c_{r_N}\}$ graded by antifield numbers Ant$[\ol
c_{r_k}]=k+2$. Let $k=-1,0$ further stand for $\ol s_A$ and $\ol
c_r$, respectively. We assume that
the BGDA $\ol\cP^*_\infty\{N\}$ (\ref{v91}) is provided with a
nilpotent graded derivation
\mar{v92,'}\ben
&&\dl_N=\dl_0 + \op\sum_{1\leq k\leq N}\rdr^{r_k} \Delta_{r_k},
\label{v92}\\
&& \Delta_{r_k}=G_{r_k} + h_{r_k}= \op\sum_{0\leq|\La|}
\Delta_{r_k}^{r_{k-1},\La}\ol c_{\La r_{k-1}} + \op\sum_{0\leq
|\Si|, |\Xi|}(h_{r_k}^{(r_{k-2},\Si)(A,\Xi)}\ol c_{\Si
r_{k-2}}\ol s_{\Xi A}+...), \label{v92'}
\een
of antifield number -1, and that
the module $\ol\cP^{0,n}_\infty\{N\}_{\leq N+3}$ of graded
densities of antifield number Ant$[\f]\leq N+3$ is split into the
$(N+1)$-exact chain complex
\mar{v94}\ben
&&0\lto \im \ol\dl \llr^{\ol\dl} \cP^{0,n}_\infty[\ol Y^*;F;Y;\ol
F^*]_1\llr^{\dl_0} \ol\cP^{0,n}_\infty\{0\}_2\llr^{\dl_1}
\ol\cP^{0,n}_\infty\{1\}_3\cdots
\label{v94}\\
&& \qquad
 \llr^{\dl_{N-1}} \ol\cP^{0,n}_\infty\{N-1\}_{N+1}
\llr^{\dl_N} \ol\cP^{0,n}_\infty\{N\}_{N+2}\llr^{\dl_N}
\ol\cP^{0,n}_\infty\{N\}_{N+3}, \nonumber
\een
which satisfies the following $(N+1)$-homology regularity condition.

\begin{defi} \label{v155} \mar{v155} One says that the complex (\ref{v94})
obeys the $(N+1)$-homology regularity condition if any
$\dl_{k<N-1}$-cycle $\f\in \ol\cP_\infty^{0,n}\{k\}_{k+3}\subset
\ol\cP_\infty^{0,n}\{k+1\}_{k+3}$ is a $\dl_{k+1}$-boundary.
\end{defi}

Note that
the $(N+1)$-exactness of the complex (\ref{v94}) implies that any
$\dl_{k<N-1}$-cycle $\f\in \ol\cP_\infty^{0,n}\{k\}_{k+3}$, $k<N$,
is a $\dl_{k+2}$-boundary, but not necessary a
$\dl_{k+1}$-boundary.

If $N=1$, the complex $\ol\cP^{0,n}_\infty\{1\}_{\leq 4}$
(\ref{v94}) restarts the complex (\ref{v87}). 
Therefore, we agree to call $\dl_N$
(\ref{v92}) the $N$-stage Koszul--Tate differential. Its
nilpotency implies the complete Noether identities (\ref{v64}),
the first-stage Noether identities (\ref{v82}), and the equalities
\mar{v93}\beq
\op\sum_{0\leq|\La|} \Delta_{r_k}^{r_{k-1},\La}d_\La
(\op\sum_{0\leq|\Si|} \Delta_{r_{k-1}}^{r_{k-2},\Si}\ol c_{\Si
r_{k-2}}) + \ol\dl(\op\sum_{0\leq |\Si|,
|\Xi|}h_{r_k}^{(r_{k-2},\Si)(A,\Xi)}\ol c_{\Si r_{k-2}}\ol
s_{\Xi A})=0  \label{v93}
\eeq
for $k=2,\ldots,N$. One can think of the equalities (\ref{v93}) as
being complete $k$-stage Noether identities because of their
properties which we will justify in the case of $k=N+1$.
Accordingly, $\Delta_{r_k}$ (\ref{v92'}) are said to be the
$k$-stage Noether operators.

Let us consider the $(N+2)$-homology of the complex (\ref{v94}). A
generic $(N+2)$-chain $\Phi\in \ol\cP^{0,n}_\infty\{N\}_{N+2}$
takes the form
\mar{v156}\beq
\Phi= G + H= \op\sum_{0\leq|\La|} G^{r_N,\La}\ol c_{\La r_N}\om +
\op\sum_{0\leq |\Si|, |\Xi|}(H^{(A,\Xi)(r_{N-1},\Si)}\ol s_{\Xi
A}\ol c_{\Si r_{N-1}}+...)\om. \label{v156}
\eeq
Let it be a cycle. The cycle condition $\dl_N\Phi=0$ implies the
equality
\mar{v145}\beq
\op\sum_{0\leq|\La|} G^{r_N,\La}d_\La (\op\sum_{0\leq|\Si|}
\Delta_{r_N}^{r_{N-1},\Si}\ol c_{\Si r_{N-1}}) +
\ol\dl(\op\sum_{0\leq |\Si|, |\Xi|}H^{(A,\Xi)(r_{N-1},\Si)}\ol
s_{\Xi A}\ol c_{\Si r_{N-1}})=0.  \label{v145}
\eeq
One can think of this equality as being the reduction condition on
the $N$-stage Noether operators (\ref{v92'}). Conversely, let
\be
\Phi= \op\sum_{0\leq|\La|} G^{r_N,\La}\ol c_{\La r_N}\om \in
\ol\cP^{0,n}_\infty\{N\}_{N+2}
\ee
be a graded density such that the reduction condition (\ref{v145})
holds. Then this reduction condition can be extended to a cycle
one as follows. It is brought into the form
\be
&& \dl_N(\op\sum_{0\leq|\La|}  G^{r_N,\La}\ol c_{\La r_N} +
\op\sum_{0\leq |\Si|, |\Xi|}H^{(A,\Xi)(r_{N-1},\Si)}\ol
s_{\Xi A}\ol c_{\Si r_{N-1}})=\\
&& \qquad  -\op\sum_{0\leq|\La|} G^{r_N,\La}d_\La h_{r_N}
+\op\sum_{0\leq |\Si|, |\Xi|}H^{(A,\Xi)(r_{N-1},\Si)}\ol s_{\Xi
A}d_\Si \Delta_{r_{N-1}}.
\ee
A glance at the expression (\ref{v92'}) shows that the term in the
right-hand side of this equality belongs to
$\ol\cP^{0,n}_\infty\{N-2\}_{N+1}$. It is a $\dl_{N-2}$-cycle and,
consequently, a $\dl_{N-1}$-boundary $\dl_{N-1}\Psi$ in accordance
with the $(N+1)$-homology regularity condition. Then the reduction
condition (\ref{v145}) is a $\ol c_{\Si r_{N-1}}$-dependent part
of the cycle condition
\be
\dl_N(\op\sum_{0\leq|\La|} && G^{r_N,\La}\ol c_{\La r_N} +
\op\sum_{0\leq |\Si|, |\Xi|}H^{(A,\Xi)(r_{N-1},\Si)}\ol s_{\Xi
A}\ol c_{\Si r_{N-1}} -\Psi)=0,
\ee
but $\dl_N\Psi$ does not make a contribution to this reduction
condition.

Being a cycle condition, the reduction condition (\ref{v145}) is
trivial either if a cycle $\Phi$ (\ref{v156}) is a
$\dl_N$-boundary or its summand $G$ is $\ol\dl$-exact. Then
a degenerate
 Lagrangian system is said to be $(N+1)$-stage
reducible if there exist non-trivial reduction conditions
(\ref{v145}), called the $(N+1)$-stage Noether identities.

\begin{theo} \label{v163} \mar{v163}
(i) The $(N+1)$-stage Noether identities can be identified to
non-trivial elements of the homology $H_{N+2}(\dl_N)$ of the
complex (\ref{v94}) iff this homology obeys the $(N+2)$-homology
regularity condition. (ii) If the homology $H_{N+2}(\dl_N)$ is
finitely generated, the complex (\ref{v94})
admits an $(N+2)$-exact extension.
\end{theo}

\begin{proof}
(i) The $(N+2)$-homology regularity condition implies that any
$\dl_{N-1}$-cycle $\Phi\in \ol\cP_\infty^{0,n}\{N-1\}_{N+2}\subset
\ol\cP_\infty^{0,n}\{N\}_{N+2}$ is a $\dl_N$-boundary. Therefore,
if $\Phi$ (\ref{v156}) is a representative of a non-trivial
element of $H_{N+2}(\dl_N)$, its summand $G$ linear in $\ol c_{\La
r_N}$ does not vanish. Moreover, it is not a $\ol\dl$-boundary.
Indeed, if $G=\ol\dl \Psi$, then
\mar{v172}\beq
\Phi=\dl_N\Psi +(\ol \dl-\dl_N)\Psi + H. \label{v172}
\eeq
The cycle condition takes the form
\be
\dl_N\Phi=\dl_{N-1}((\ol\dl-\dl_N)\Psi + H)=0.
\ee
Hence, $(\ol \dl-\dl_N)\Psi + H$ is $\dl_N$-exact since any
$\dl_{N-1}$-cycle $\f\in \ol\cP_\infty^{0,n}\{N-1\}_{N+2}$ is a
$\dl_N$-boundary. Consequently, $\Phi$ (\ref{v172}) is a boundary.
If the $(N+2)$-homology regularity condition does not hold,
trivial reduction conditions (\ref{v145}) also come from
non-trivial elements of the homology $H_{N+2}(\dl_N)$. (ii) Let
the $(N+1)$-stage Noether identities be finitely generated.
Namely, there exist elements $\Delta_{(N+1)}\in H_{N+2}(\dl_N)$
making up a $\Bbb Z_2$-graded projective $C^\infty(X)$-module
$\cC_{(N+1)}$ of finite rank which is isomorphic to the module of
sections of the product $\ol V^*_{N+1}\op\times_X \ol E^*_{N+1}$
of the density-duals of some vector bundles $V_{N+1}\to X$ and
$E_{N+1}\to X$. Let $\{\Delta_{r_{N+1}}\}$ be local bases for this
$C^\infty(X)$-module. Then any element $\Phi\in H_{N+2}(\dl_N)$
factorizes
\mar{v160,1}\ben
&& \Phi= \op\sum_{0\leq|\Xi|} \Phi^{r_{N+1},\Xi} d_\Xi
\Delta_{r_{N+1}}\om, \qquad \Phi^{r_{N+1},\Xi}\in
\cS^0_\infty[F;Y], \label{v160}\\
&&\Delta_{r_{N+1}}=G_{r_{N+1}}+ h_{r_{N+1}}=\op\sum_{0\leq|\La|}
\Delta_{r_{N+1}}^{r_N,\La}\ol c_{\La r_N} + h_{r_{N+1}},
\label{v161}
\een
via elements of $\cC_{(N+1)}$. Clearly, this factorization is
independent of specification of local bases
$\{\Delta_{r_{N+1}}\}$. Let us extend the BGDA
$\ol\cP^*_\infty\{N\}$ (\ref{v91}) to the  BGDA
$\ol\cP^*_\infty\{N+1\}$ possessing the local basis
$\{s^A,\ol s_A, \ol c_r, \ol c_{r_1}, \ldots, \ol c_{r_N}, \ol
c_{r_{N+1}}\}$ where ${\rm Ant}[\ol c_{r_{N+1}}]=N+3$ and $[\ol
c_{r_{N+1}}]=([\Delta_{r_{N+1}}]+1){\rm mod}\,2$.
It is provided with the nilpotent graded derivation
$\dl_{N+1}=\dl_N + \rdr^{r_{N+1}} \Delta_{r_{N+1}}$
of antifield number -1. With this graded derivation, the module
$\ol\cP^{0,n}_\infty\{N+1\}_{\leq N+4}$ of graded densities of
antifield number Ant$[\f]\leq N+4$ is split into the chain complex
\mar{v171}\ben
&&0\lto \im \ol\dl \llr^{\ol\dl} \cP^{0,n}_\infty[\ol Y^*;F;Y;\ol
F^*]_1\llr^{\dl_0} \ol\cP^{0,n}_\infty\{0\}_2\llr^{\dl_1}
\ol\cP^{0,n}_\infty\{1\}_3\cdots
 \label{v171}\\
&& \quad \llr^{\dl_{N-1}} \ol\cP^{0,n}_\infty\{N-1\}_{N+1}
 \llr^{\dl_N} \ol\cP^{0,n}_\infty\{N\}_{N+2}\llr^{\dl_{N+1}}
\ol\cP^{0,n}_\infty\{N+1\}_{N+3}\llr^{\dl_{N+1}}
\ol\cP^{0,n}_\infty\{N+1\}_{N+4}. \nonumber
\een
It is readily observed that this complex is $(N+2)$-exact. In this
case, the $(N+1)$-stage Noether identities (\ref{v145}) come from
the complete $(N+1)$-stage Noether identities
\mar{v162}\beq
 \op\sum_{0\leq|\La|} \Delta_{r_{N+1}}^{r_N,\La} d_\La \Delta_{r_N}\om
+\ol\dl h_{r_{N+1}}\om =0, \label{v162}
\eeq
which are reproduced as the nilpotency conditions of the graded
derivation $\dl_{N+1}$.
\end{proof}

The iteration procedure based on Theorem \ref{v163} may be
infinite. We restrict our consideration to the case of a finitely
($N$-stage) reducible Lagrangian system possessing the finite
$(N+2)$-exact Koszul--Tate complex
\mar{w3,4}\ben
&&0\lto \im \ol\dl \llr^{\ol\dl} \cP^{0,n}_\infty[\ol Y^*;F;Y;\ol
F^*]_1\llr^{\dl_0} \ol\cP^{0,n}_\infty\{0\}_2\llr^{\dl_1}
\ol\cP^{0,n}_\infty\{1\}_3\cdots
\label{w3}\\
&& \qquad
 \llr^{\dl_{N-1}} \ol\cP^{0,n}_\infty\{N-1\}_{N+1}
\llr^{\dl_N} \ol\cP^{0,n}_\infty\{N\}_{N+2}\llr^{\dl_N}
\ol\cP^{0,n}_\infty\{N\}_{N+3}, \nonumber\\
&&\dl_N=\rdr^A \cE_A + \op\sum_{0\leq|\La|} \rdr^r
\Delta_r^{A,\La}\ol s_{\La A} + \op\sum_{1\leq k\leq N}\rdr^{r_k}
\Delta_{r_k}, \label{w4}
\een
where $\Delta_{r_k}$ are the $k$-stage Noether operators
(\ref{v92'}) which obey the Noether identities (\ref{v64}),
first-stage Noether identities (\ref{v82}) and $k$-stage Noether
identities (\ref{v93}), $k=2,\ldots,N$. 
Forthcoming Theorem
\ref{w35} associates to the
Koszul--Tate complex (\ref{w3}) the sequence (\ref{w36}), graded in
ghosts, whose ascent operator $\up_e$ (\ref{w108}) provides
the gauge and higher-stage gauge supersymmetries of an original
graded Lagrangian.

\section{The Noether second theorem}

Given the BGDA $\ol\cP^*_\infty\{N\}$ (\ref{v91}), let us consider
the BGDA
\mar{w5}\beq
\cP^*_\infty\{N\}=\cP^*_\infty[V_N\cdots V_1V;F;Y;EE_1\cdots E_N]
\label{w5}
\eeq
with the local basis $\{s^A, c^r, c^{r_1}, \ldots, c^{r_N}\}$ and the
BGDA
\mar{w6}\beq
P^*_\infty\{N\}=\cP^*_\infty[\ol E^*_N\cdots\ol E^*_1\ol E^*\ol
Y^*V_N\cdots V_1V;F;Y;EE_1\cdots E_N\ol F^*\ol V^*\ol
V^*_1\cdots\ol V_N^*] \label{w6}
\eeq
with the local basis
\mar{w7}\beq
\{s^A, c^r, c^{r_1}, \ldots, c^{r_N},\ol s_A,\ol c_r, \ol c_{r_1},
\ldots, \ol c_{r_N}\},\label{w7}
\eeq
where $[c^{r_k}]=([\ol c_{r_k}]+1){\rm mod}\,2$ and
Ant$[c^{r_k}]=-(k+1)$. We call $c^{r_k}$, $k=0,\ldots,N$, the ghosts of
ghost number gh$[c^{r_k}]=k+1$. Clearly, the BGDAs
$\ol\cP^*_\infty\{N\}$ (\ref{v91}) and $\cP^*_\infty\{N\}$
(\ref{w5}) are subalgebras of the BGDA $P^*_\infty\{N\}$
(\ref{w6}). The Koszul--Tate differential $\dl_N$ (\ref{w4}) is
naturally extended to a graded derivation of the BGDA
$P^*_\infty\{N\}$ (\ref{w6}).

\begin{theo} \label{w35} \mar{w35}
With the Koszul--Tate complex (\ref{w3}) of antifields, the graded
commutative ring $\cP_\infty^0\{N\}\subset \cP_\infty^*\{N\}$
(\ref{w5}) of ghosts is split into the sequence
\mar{w36,108}\ben
&& 0\to \cS^0_\infty[F;Y]\ar^{u_e} \cP^0_\infty\{N\}_1\ar^{u_e}
\cP^0_\infty\{N\}_2\ar^{u_e}\cdots, \label{w36} \\
&& u_e=u + \op\sum_{1\leq k\leq N} u_{(k)}, 
\label{w108}
\een
where $u$ (\ref{w33}), $u_{(1)}$ (\ref{w38'}) and $u_{(k)}$ (\ref{w38}),
$k=2,\ldots, N$,  are the operators of gauge and higher-stage gauge
supersymmetries of an original graded Lagrangian. 
\end{theo}

\begin{proof}
Let us extend an original graded Lagrangian $L$ to
the even graded density
\mar{w8}\beq
L_e=\cL_e\om=L+L_1=L + \op\sum_{0\leq k\leq N} c^{r_k}\Delta_{r_k}\om=L
+\dl_N( \op\sum_{0\leq k\leq N} c^{r_k}\ol c_{r_k}\om), \label{w8}
\eeq
whose summand $L_1$ is linear in ghosts and  
possesses the zero antifield number. 
It is readily observed that
$\dl_N(L_e)=0$, i.e., $\dl_N$ is a variational supersymmetry of
the graded Lagrangian $L_e$ (\ref{w8}). It follows that
\mar{w16}\ben
 && [\frac{\op\dl^\lto \cL_e}{\dl \ol
s_A}\cE_A +\op\sum_{0\leq k\leq N} \frac{\op\dl^\lto \cL_e}{\dl \ol
c_{r_k}}\Delta_{r_k}]\om = [\frac{\op\dl^\lto \cL_e}{\dl \ol
s_A}\cE_A +\op\sum_{0\leq k\leq N} \frac{\op\dl^\lto \cL_e}{\dl \ol
c_{r_k}} \frac{\dl \cL_e}{\dl
c^{r_k}}]\om= \label{w16}\\
&& \qquad [\up^A\cE_A + \op\sum_{0\leq k\leq N}\up^{r_k}\frac{\dl
\cL_e}{\dl c^{r_k}}]\om= d_H\si, \nonumber\\
&& \up^A= \frac{\op\dl^\lto \cL_e}{\dl \ol
s_A}=u^A+w^A =\op\sum_{0\leq|\La|} c^r_\La\eta(\Delta^A_r)^\La +
\op\sum_{i>0}\op\sum_{0\leq|\La|}
c^{r_i}_\La\eta(\op\dr^\lto{}^A(h_{r_i}))^\La, \nonumber\\
&& \up^{r_k}=\frac{\op\dl^\lto \cL_e}{\dl \ol
c_{r_k}} =u^{r_k}+ w^{r_k}= \op\sum_{0\leq|\La|}
c^{r_{k+1}}_\La\eta(\Delta^{r_k}_{r_{k+1}})^\La +\op\sum_{i>k+1}
\op\sum_{0\leq|\La|} c^{r_i}_\La\eta(\op\dr^\lto{}^{r_k}(h_{r_i}))^\La,
\nonumber
\een
(see the formulas (\ref{0606b}) -- (\ref{0606c})). The equality
(\ref{w16}) falls into the set of equalities
\mar{w19,',20}\ben
&& \frac{\op\dl^\lto (c^r\Delta_r)}{\dl \ol s_A}\cE_A\om
=u^A\cE_A\om=d_H\si_0, \label{w19}\\
&&  [\frac{\op\dl^\lto (c^{r_1}\Delta_{r_1})}{\dl \ol s_A}\cE_A
+\frac{\op\dl^\lto (c^{r_1}\Delta_{r_1})}{\dl \ol
c_r}\Delta_r]\om= d_H\si_1, \label{w19'}\\
&&  [\frac{\op\dl^\lto (c^{r_i}\Delta_{r_i})}{\dl \ol s_A}\cE_A
+\op\sum_{k<i} \frac{\op\dl^\lto (c^{r_i}\Delta_{r_i})}{\dl \ol
c_{r_k}}\Delta_{r_k}]\om= d_H\si_i, \qquad
i=2,\ldots,N,\label{w20}
\een
with respect to the polynomial degree in ghosts.
A glance at the equality (\ref{w19}) shows that, by virtue of the
first variational formula (\ref{g107}), the graded derivation
\mar{w33}\beq
u= u^A\frac{\dr}{\dr s^A}, \qquad u^A
=\op\sum_{0\leq|\La|} c^r_\La\eta(\Delta^A_r)^\La, \label{w33}
\eeq
is a variational supersymmetry of an original graded Lagrangian $L$. 
This variational supersymmetry is parameterized by
ghosts $c^r$. Therefore, one can think of it as being a gauge
supersymmetry of $L$ \cite{jmp05,cmp04}.
The equality (\ref{w19'}) takes the form
\be
&&[\frac{\op\dl^\lto}{\dl \ol s_A}(c^{r_1}h_{r_1}^{(B,\Si)(A,\Xi)}
\ol s_{\Si B}\ol s_{\Xi A})\cE_A + \frac{\op\dl^\lto}{\dl \ol
c_r}(c^{r_1}\op\sum_{0\leq|\Si|}\Delta_{r_1}^{r,\Si}\ol c_{\Si
r})\op\sum_{0\leq|\Xi|} \Delta_r^{B,\Xi}\ol s_{\Xi B}]\om= \\
&& \qquad [\op\sum_{0\leq|\Xi|}
(-1)^{|\Xi|}d_\Xi(c^{r_1}\op\sum_{0\leq|\Si|} 2h_{r_1}^{(B,\Si)(A,\Xi)}
\ol s_{\Si B})\cE_A + u^r\op\sum_{0\leq|\Xi|}
\Delta_r^{B,\Xi}\ol s_{\Xi B}]\om= d_H\si'_1.
\ee
Using the relation (\ref{0606a}), we obtain
\be
[\op\sum_{0\leq|\Xi|} c^{r_1}\op\sum_{0\leq|\Si|} 2h_{r_1}^{(B,\Si)(A,\Xi)} \ol
s_{\Si B} d_\Xi\cE_A + u^r\op\sum_{0\leq|\Xi|} \Delta_r^{B,\Xi}\ol s_{\Xi
B}]\om= d_H\si_1.
\ee
The variational derivative of the both sides of this
equality with respect to the antifield $\ol s_B$ leads to the relation
\be
\op\sum_{0\leq|\Si|} \eta(h_{r_1}^{(B)(A,\Xi)})^\Si d_\Si(2c^{r_1}
d_\Xi\cE_A) + \op\sum_{0\leq|\Si|} u^r_\Si\eta (\Delta^B_r)^\Si=0,
\ee
which is brought into the form
\mar{w34'}\beq
\op\sum_{0\leq|\Si|} d_\Si u^r\frac{\dr}{\dr c^r_\Si} u^B=\ol\dl(\al^B), 
\qquad \al^B = -\op\sum_{0\leq|\Si|}
\eta(2h_{r_1}^{(B)(A,\Xi)})^\Si d_\Si(c^{r_1} \ol s_{\Xi A}).
\label{w34'}
\eeq
Therefore, the odd graded derivation
\mar{w38'}\beq
u_{(1)}= u^r\frac{\dr}{\dr c^r}, \qquad u^r=\op\sum_{0\leq|\La|}
c^{r_1}_\La\eta(\Delta^r_{r_1})^\La, \label{w38'}
\eeq
is the first-stage gauge supersymmetry of a
reducible Lagrangian  system \cite{jmp05}.
Every equality (\ref{w20}) is split into a set of equalities with
respect to the polynomial degree in antifields. Let us consider the
one, linear in antifields $\ol c_{r_{i-2}}$ and their jets. We have
\be
&& [\frac{\op\dl^\lto}{\dl \ol
s_A}(c^{r_i}\op\sum_{0\leq|\Si|,|\Xi|}h_{r_i}^{(r_{i-2},\Si)(A,\Xi)} \ol
c_{\Si r_{i-2}}\ol s_{\Xi A})\cE_A + \\
&& \qquad \frac{\op\dl^\lto}{\dl \ol
c_{r_{i-1}}}(c^{r_i}\op\sum_{0\leq|\Si|}\Delta_{r_i}^{r'_{i-1},\Si}\ol
c_{\Si r'_{i-1}})\op\sum_{0\leq|\Xi|} \Delta_{r_{i-1}}^{r_{i-2},\Xi}\ol
c_{\Xi r_{i-2}}]\om= d_H\si_i.
\ee
It is brought into the form
\be
[\op\sum_{0\leq|\Xi|} (-1)^{|\Xi|}d_\Xi(c^{r_i}\op\sum_{0\leq|\Si|}
h_{r_i}^{(r_{i-2},\Si)(A,\Xi)} \ol c_{\Si r_{i-2}})\cE_A +
u^{r_{i-1}}\op\sum_{0\leq|\Xi|} \Delta_{r_{i-1}}^{r_{i-2},\Xi}\ol c_{\Xi
r_{i-2}}]\om= d_H\si_i.
\ee
Using the relation (\ref{0606a}), we obtain
\be
[\op\sum_{0\leq|\Xi|} c^{r_i}\op\sum_{0\leq|\Si|} h_{r_i}^{(r_{i-2},\Si)(A,\Xi)} \ol
c_{\Si r_{i-2}} d_\Xi\cE_A + u^{r_{i-1}}\op\sum_{0\leq|\Xi|}
\Delta_{r_{i-1}}^{r_{i-2},\Xi}\ol c_{\Xi r_{i-2}}]\om= d_H\si'_i.
\ee
The variational derivative of the both sides of this equality
with respect to the antifield $\ol c_{r_{i-2}}$ leads to the relation
\be
\op\sum_{0\leq|\Si|} \eta(h_{r_i}^{(r_{i-2})(A,\Xi)})^\Si d_\Si(c^{r_i}
d_\Xi\cE_A) + \op\sum_{0\leq|\Si|} u^{r_{i-1}}_\Si\eta
(\Delta^{r_{i-2}}_{r_{i-1}})^\Si=0,
\ee
which takes the form
\mar{w34}\beq
\op\sum_{0\leq|\Si|} d_\Si u^{r_{i-1}}\frac{\dr}{\dr c^{r_{i-1}}_\Si} u^{r_{i-2}}
=\ol\dl(\al^{r_{i-2}}),
\qquad \al^{r_{i-2}} = -\op\sum_{0\leq|\Si|} \eta(h_{r_i}^{(r_{i-2})(A,\Xi)})^\Si
d_\Si(c^{r_i} \ol s_{\Xi A}). \label{w34}
\eeq
Therefore, the odd graded derivations
\mar{w38}\beq
u_{(k)}= u^{r_{k-1}}\frac{\dr}{\dr c^{r_{k-1}}}, \qquad
u^{r_{k-1}}=\op\sum_{0\leq|\La|}
c^{r_k}_\La\eta(\Delta^{r_{k-1}}_{r_k})^\La, \qquad k=2,\ldots,N,
\label{w38}
\eeq
are the $k$-stage gauge supersymmetries
\cite{jmp05}.
The graded derivations $u$ (\ref{w33}), $u_{(1)}$ (\ref{w38'}), 
$u_{(k)}$ (\ref{w38})  are assembled into the ascent operator
(\ref{w108}) of ghost number 1, that we agree to call
the total gauge operator. It provides the sequence (\ref{w36}).
\end{proof}

The total gauge operator (\ref{w108}) need not be
nilpotent even on the shell. 
We say that gauge and higher-stage gauge supersymmetries of a 
Lagrangian system form an algebra on the shell if the graded
derivation (\ref{w108}) can be extended to a graded derivation
$\up^0$ of ghost number 1 by means of terms of higher 
polynomial degree in ghosts such that $\up^0$ is nilpotent on the shell.
Namely, we have
\mar{w109}\beq
\up^0=u_e+ \xi= u^A\dr_A + \op\sum_{1\leq k\leq N}(u^{r_{k-1}}
+\xi^{r_{k-1}})\dr_{r_{k-1}}, \label{w109}
\eeq 
where all the coefficients $\xi^{r_{k-1}}$ are at least 
quadratic in ghosts and $(\up^0\circ\up^0)(f)$ is $\ol\dl$-exact
for any graded function 
$f\in \cP^0_\infty\{N\}\subset P^0_\infty\{N\}$. This nilpotency 
condition falls into a set of equalities with respect to
the polynomial degree in ghosts. Let us write the first and second of 
them involving the coefficients $\xi_2^{r_{k-1}}$ quadratic in ghosts.
We have
\mar{w110,3}\ben
&& \op\sum_{0\leq|\Si|} d_\Si u^r\dr^\Si_r u^B=\ol\dl(\al^B_1), \qquad
\op\sum_{0\leq|\Si|} d_\Si u^{r_{k-1}}\dr_{r_{k-1}}^\Si u^{r_{k-2}}
=\ol\dl(\al^{r_{k-2}}_1), \quad 2\leq k\leq N, \label{w110}\\
&&  \op\sum_{0\leq|\Si|}[d_\Si u^A\dr^\Si_A u^B 
+d_\Si\xi^r_2\dr^\Si_r u^B]=\ol\dl(\al^B_2), 
\label{w111} \\
&& \op\sum_{0\leq|\Si|}[d_\Si u^A\dr^\Si_A u^{r_{k-1}} +
d_\Si\xi^{r_k}_2\dr^\Si_{r_k} u^{r_{k-1}} +
d_\Si
u^{r'_{k-1}}\dr^\Si_{r'_{k-1}}\xi^{r_{k-1}}_2]= \ol\dl(\al^{r_{k-1}}_2),
 \label{w112}\\
&& \xi_2^r=\xi^{r,\La,\Si}_{r',r''} c^{r'}_\La c^{r''}_\Si, 
\qquad  \xi_2^{r_k}=\xi^{r_k,\La,\Si}_{r,r'_k} c^r_\La c^{r'_k}_\Si, \qquad 
2\leq k\leq N. \label{w113}
\een
The equalities (\ref{w110}) reproduce the relations (\ref{w34'}) and 
(\ref{w34}) in Theorem \ref{w35}. The equalities (\ref{w111}) -- (\ref{w112})
provide the generalized
commutation relations on the shell between gauge and higher-stage gauge 
supersymmetries, and one  can think of the coefficients $\xi_2$ (\ref{w113})
as being {\it sui generis} generalized structure functions
\cite{jmp05,fulp02}.

Note that the total gauge operator in an irreducible gauge
theory  is the operator of
infinitesimal gauge transformations whose parameter functions  are
replaced with the ghosts. Its nilpotent extension is the familiar
BRST operator \cite{ijgmmp05,cmp04}.
For instance, let $P\to X$ be a principal bundle  with a structure Lie
group
$G$, whose Lie algebra possesses the basis $\{e_r\}$ and the structure
constants
$c^r_{pq}$. Let us consider a gauge theory of principal connections on
$P$. It is an irreducible degenerate Lagrangian
system. Principal connections on
$P$ are represented by sections of the quotient
$C=J^1P/G$, called the bundle of principal connections. 
It is an affine bundle coordinated by $(x^\la, a^r_\la)$ such
that, given a section $A$ of $C\to X$, its components
$A^r_\la=a^r_\la\circ A$ are coefficients of the local
connection form (i.e., gauge potentials). 
Infinitesimal generators of one-parameter groups of automorphisms
of a principal bundle $P$ are $G$-invariant projectable vector
fields on $P$. They are associated to sections of the vector
bundle $T_GP=TP/G$ provided with the bundle
coordinates $(x^\la,\dot x^\la,\xi^r)$ with respect to the fiber
bases $\{\dr_\la, e_r\}$ for $T_GP$. The form an algebra. Given sections
\mar{0652}\beq
Êu=u^\la\dr_\la
+u^r e_r, \qquad v=v^\la\dr_\la +v^r e_r, \label{0652}
\eeq
of $T_GP\to X$, their bracket reads
\be
[u,v]=(u^\m\dr_\m v^\la -v^\m\dr_\m u^\la)\dr_\la +(u^\la\dr_\la
v^r - v^\la\dr_\la u^r +c^r_{pq}u^pv^q)e_r. 
\ee
Any section $u$ (\ref{0652}) of the vector bundle $T_GP\to X$ yields the
vector field
\be
u_C=u^\la\dr_\la +(c^r_{pq}a^p_\la u^q +\dr_\la u^r- a^r_\m\dr_\la
u^\m)\dr^\la_r 
\ee
on the bundle of principal connections $C$, i.e., the 
infinitesimal gauge transformation with the parameter functions
$u^\la$ and $u^r$. Taking its vertical part and replacing parameter
functions with the corresponding ghosts
$c^\la$ and $c^r$, we obtain the total gauge operator
\be
\up_e= (c^r_{pq}a^p_\la c^q + c^r_\la
-a^r_\m c^\m_\la-c^\m a_{\m\la}^r)\frac{\dr}{\dr a_\la^r}. 
\ee
Its nilpotent extension is the BRST operator
\be
\up^0= \up_e +(-\frac12c^r_{pq}c^pc^q -c^\m c^r_\m)\frac{\dr}{\dr c^r}
+ c^\la_\m c^\m\frac{\dr}{\dr c^\la}. 
\ee

\section{The master equation}

The BGDA $P^*_\infty\{N\}$ (\ref{w6}) with the local basis
(\ref{w7}) exemplifies Lagrangian systems of the following particular
type.

Let $Y_0\to X$ be a fiber bundle admitting the vertical splitting
$VY_0=Y_0\times_X W$, where $W\to X$ is a vector bundle whose
density-dual is denoted by $\ol Y^*_0$. Let $Y_1\to X$ be a vector
bundle and $\ol Y_1^*$ its density-dual. We consider the BGDA
$\cP^*_\infty[\ol Y^*_0;Y_1;Y_0;\ol Y_1^*]$ endowed with the local
basis $\{y^a,\ol y_a\}$, where $[\ol y_a]=([y^a]+1){\rm mod}\,2$.
Let us call $y^a$ and $\ol y_a$
the fields and antifields, respectively. Then one can associate to any
graded Lagrangian
\mar{w40}\beq
\gL\om\in \cP^{0,n}_\infty[\ol Y^*_0;Y_1;Y_0;\ol Y_1^*]
\label{w40}
\eeq
the odd graded derivations
\mar{w37}\beq
\up=\op\cE^\lto{}^a\dr_a=\frac{\op\dl^\lto \gL}{\dl \ol y_a}
\frac{\dr}{\dr y^a},
\qquad \ol\up= \rdr^a\cE_a=\frac{\op\dr^\lto}{\dr \ol y_a}\frac{\dl
\gL}{\dl y^a} \label{w37}
\eeq
of the BGDA $\cP^*_\infty[\ol Y^*_0;Y_1;Y_0;\ol Y_1^*]$.

\begin{prop} \label{w39} \mar{w39} The following conditions are
equivalent:

(i) the graded derivation $\up$ (\ref{w37}) is a variational 
supersymmetry of a Lagrangian $\gL\om$ (\ref{w40})

(ii) the graded derivation $\ol\up$ (\ref{w37}) is a variational 
supersymmetry of $\gL\om$ (\ref{w40}),

(iii) the composition $(\up-\ol\up)\circ(\up +\ol\up)$ acting on 
even graded functions $f\in 
\cP^0_\infty[\ol Y^*_0;Y_1;Y_0;\ol Y_1^*]$ (or, equivalently,
$(\up+\ol\up)\circ(\up -\ol\up)$ acting on 
the odd ones) vanishes.
\end{prop}

\begin{proof} By virtue of the first variational formula
(\ref{g107}) (see also Remark \ref{w74}), the 
conditions (i) and (ii) are equivalent to the equality
\mar{w44}\beq
\op\cE^\lto{}^a\cE_a\om=\frac{\op\dl^\lto \gL}{\dl \ol
y_a}\frac{\dl \gL}{\dl y^a}\om =d_H\si. \label{w44}
\eeq
In accordance with Theorem \ref{v11} and Proposition \ref{w46},
the equality (\ref{w44}) is equivalent to the condition that the
odd graded density $\op\cE^\lto{}^a\cE_a\om$ is variationally
trivial. For convenience, let us replace the right variational
derivatives $\op\cE^\lto{}^a$ in the equality (\ref{w44}) with the
left ones $(-1)^{[a]+1}\cE^a$. We obtain
\mar{w91}\beq
\op\sum_a (-1)^{[a]}\cE^a\cE_a\om=d_H\si. \label{w91}
\eeq
The variational operator acting on this
relation leads to the equalities
\be
&& \op\sum_{0\leq|\La|}(-1)^{[a]+|\La|}d_\La(\dr^\La_b(\cE^a\cE_a))=
\op\sum_{0\leq|\La|}(-1)^{[a]}[\eta(\dr_b\cE^a)^\La\cE_{\La a} +
\eta(\dr_b\cE_a)^\La\cE^a_\La)]=0, \\
&& \op\sum_{0\leq|\La|}(-1)^{[a]+|\La|}d_\La(\dr^{\La b}(\cE^a\cE_a)) =
\op\sum_{0\leq|\La|}(-1)^{[a]}[\eta(\dr^b\cE^a)^\La\cE_{\La a} +
\eta(\dr^b\cE_a)\cE^a_\La] = 0.
\ee
Due to the formulas (\ref{w51}), these equalities are
brought into the form
\mar{w92,3}\ben
&& \op\sum_{0\leq|\La|}(-1)^{[a]}[(-1)^{[b]([a]+1)}\dr^{\La a}\cE_b\cE_{\La
a} + (-1)^{[b][a]}\dr_a^\La\cE_b\cE^a_\La]=0, \label{w92}\\
&& \op\sum_{0\leq|\La|}(-1)^{[a]}[(-1)^{([b]+1)([a]+1)}\dr^{\La
a}\cE^b\cE_{\La a} + (-1)^{([b]+1)[a]}\dr_a^\La\cE^b\cE^a_\La]=0,
\label{w93}
\een
for all $\cE_b$ and $\cE^b$. Returning to the right variational
derivatives, we obtain the relations
\mar{w55,6}\ben
&& \op\dr^\lto{}^{\La a}(\cE_b)\cE_{\La a} +
(-1)^{[b]}\op\cE^\lto{}^a_\La \dr^\La_a\cE_b=0, \label{w55}\\
&& \op\cE^\lto{}^a_\La \dr^\La_a\op\cE^\lto{}^b + (-1)^{[b]+1}
\op\dr^\lto{}^{\La a}(\op\cE^\lto{}^b)\cE_{\La a}=0. \label{w56}
\een
A direct computation shows that they are equivalent to the condition (iii). 
\end{proof}

Following the terminology of BV quantization, we say that a graded Lagrangian
(\ref{w40}) obeys the master equation (\ref{w44}). 

For instance, any variationally trivial Lagrangian $L_0\in
\cP^{0,n}_\infty[\ol Y^*_0;Y_1;Y_0;\ol Y_1^*]$ in Corollary
\ref{cmp26} satisfies  the master
equation. We say that a solution of the master equation is not
trivial if both the graded derivations (\ref{w37}) are not zero. It is
readily observed that,  if a graded Lagrangian
$\gL\om$ provides a nontrivial solution of the master equation and $L_0$
is a variationally trivial Lagrangian, its sum
$\gL\om +L_0$ is also a nontrivial solution of the master equation.

\begin{rem}
By virtue of Proposition \ref{w39}, the master equation (\ref{w44}) 
is equivalent to the 
equalities (\ref{w55}) -- (\ref{w56}). It is readily observed that 
these equalities are Noether 
identities of a Lagrangian (\ref{w40}) indexed by the variational derivatives 
$\cE_b$ and  $\op\cE^\lto{}^b$.
Rewritten with respect to the left variational derivatives,
these Noether identities take the form (\ref{w92}) -- (\ref{w93}). 
By virtue of Theorem \ref{w35}, the Noether identities
(\ref{w92}) -- (\ref{w93})  
define the gauge supersymmetry $u$ (\ref{w33}) of 
the Lagrangian (\ref{w40}) which is parameterized by 
the corresponding ghosts $c_b$, $c^b$.
 Using the formulas (\ref{w51}), one obtains
\be
&& u=\op\sum_a \op\sum_{0\leq|\La|} (-1)^{[a]}[c^b_\La(\dr^\La_b\cE^a\dr_a +
\dr^\La_b\cE_a\dr^a) +
 c_{\La b}(\dr^{\La b}\cE^a\dr_a +\dr^{\La b}\cE_a\dr^a)],\\
&& (u^a\cE_a + u_a\cE^a)\om= \op\sum_a \op\sum_{0\leq|\La|} 
(-1)^{[a]}[c^b_\La(\dr^\La_b\cE^a\cE_a +
\dr^\La_b\cE_a\cE^a) +
 c_{\La b}(\dr^{\La b}\cE^a\cE_a +\dr^{\La b}\cE_a\cE^a)]\om=\\
&& \qquad \op\sum_{0\leq|\La|} (c^b_\La\dr^\La_b + c_{\La b}\dr^{\La b})
(\op\sum_a (-1)^{[a]}\cE^a\cE_a\om)= d_H[
\op\sum_{0\leq|\La|} (c^b_\La\dr^\La_b + c_{\La b}\dr^{\La b})\si],
\ee
where the last equality results from action of the graded derivation
$c_b\dr^b+c^b\dr_b$ on the both sides of the master equation (\ref{w91}).
\end{rem}

Let us return to an original Lagrangian system
$(\cS^*_\infty[F;Y],L)$ and its extension $(P^*_\infty\{N\},L_e)$
to ghosts and antifields, together with the odd graded derivations (\ref{w37})
which read
\mar{w80,1}\ben
&& \up_e= \vt +\vt_e= \frac{\op\dl^\lto \cL_1}{\dl \ol
s_A}\frac{\dr}{\dr s^A} + \op\sum_{0\leq k\leq N}
\frac{\op\dl^\lto \cL_1}{\dl \ol
c_{r_k}}\frac{\dr}{\dr c^{r_k}}, \label{w80} \\
&& \ol\up_e=\ol\vt +\dl_N  = \frac{\rdr }{\dr \ol
s_A}\frac{\dl\cL_1}{\dl s^A} +
[\frac{\rdr }{\dr \ol
s_A}\frac{\dl\cL}{\dl s^A} +\op\sum_{0\leq k\leq N}
\frac{\rdr }{\dr \ol
c_{r_k}}\frac{\dl \cL_1}{\dl c^{r_k}}]. \label{w81}
\een

An original Lagrangian provides a trivial solution of the master equation.
It follows at once from the equality (\ref{w16}) that 
the graded Lagrangian $L_e$ (\ref{w8}) satisfies the master 
equation (\ref{w44}) iff
\mar{w72}\beq
\frac{\op\dl^\lto \cL_1}{\dl \ol
s_A}\frac{\dl\cL_1}{\dl s^A}\om= d_H\si'. \label{w72}
\eeq
If the condition (\ref{w72}) does not hold, a problem is to extend 
the graded Lagrangian $L_e$ (\ref{w8})
to a solution of the master equation
\mar{w61}\beq
L_e+L'=L+L_1+L_2+\cdots \label{w61}
\eeq
by means of terms $L_i$ of polynomial degree $i>1$ in ghosts.
They are assumed to be even of zero 
antifield number. 
Such an extension need not exists. Our goal is to investigate the 
conditions of its
existence (Theorems \ref{w120} and \ref{w130}).

Let a graded Lagrangian (\ref{w61}) be a solution
of the master equation (\ref{w44}),
which reads
\mar{w75}\beq
{\op\dl^\lto}{}^A(L_1+L')\dl_A(L+L_1+L')
+\op\sum_{0\leq k\leq N} {\op\dl^\lto}{}^{r_k}(L_1+L')
\dl_{r_k}(L_1+L')= d_H\si. \label{w75}
\eeq
As was mentioned above, such a solution is never unique,
but it is defined at least up to a
$d_H$-exact density.
The master equation (\ref{w75}) 
is decomposed into a set of equalities with respect to the polynomial
degree  in ghosts. We have
\mar{w76,8}\ben
&& {\op\dl^\lto}{}^A(L_1)\dl_AL
+\op\sum_{0\leq k\leq N} {\op\dl^\lto}{}^{r_k}(L_1)
\dl_{r_k}L_1 = d_H\si_1, \label{w76}\\
&& \op\sum_{1\leq j<i}{\op\dl^\lto}{}^A(L_j)\dl_AL_{i-j}
+ \op\sum_{1<j<i} \op\sum_{0\leq k\leq N}{\op\dl^\lto}{}^{r_k}(L_j)
\dl_{r_k}L_{i-j+1} + \label{w78}\\
&& \qquad {\op\dl^\lto}{}^A(L_i)\dl_AL
+\op\sum_{0\leq k\leq N}[ {\op\dl^\lto}{}^{r_k}(L_1)
\dl_{r_k}L_i +  {\op\dl^\lto}{}^{r_k}(L_i)
\dl_{r_k}L_1] = d_H\si_i, \quad i\geq 2. \nonumber
\een
The first one is exactly 
the equality (\ref{w16}). The others 
are brought into the form
\mar{w83,95}\ben
&& \op\sum_{1\leq j<i}{\op\dl^\lto}{}^A(L_j)\dl_AL_{i-j}
+ \op\sum_{1<j<i} \op\sum_{0\leq k\leq N}{\op\dl^\lto}{}^{r_k}(L_j)
\dl_{r_k}L_{i-j+1} + \g(L_i)=  d_H\si'_i, \quad i\geq 2, \label{w83}\\
&& \g=\dl_N +\vt_e = \frac{\rdr }{\dr \ol
s_A}\frac{\dl\cL}{\dl s^A} +\op\sum_{0\leq k\leq N}
[\frac{\rdr }{\dr \ol
c_{r_k}}\frac{\dl \cL_1}{\dl c^{r_k}}+ \frac{\op\dl^\lto \cL_1}{\dl \ol
c_{r_k}}\frac{\dr}{\dr c^{r_k}}]. \label{w95}
\een
It is readily observed  that 
a graded Lagrangian (\ref{w61}) obeys the 
master equation (\ref{w75}) iff its term $L_2$ is a solution of the equation
(\ref{w83}), $i=2$, the term $L_3$ satisfies the equation (\ref{w83}), 
$i=3$, and so on. Since $\g$ (\ref{w95}) vanishes on
functions $f\in \cS^0_\infty[F;Y]$, each equation (\ref{w83}) reduces to 
a system of linear algebraic equations with coefficients in the ring 
$\cS^0_\infty[F;Y]$ whose homogeneous part is given by the operator
$\g$ (\ref{w95}). Because this operator is not invertible as a rule,
a solution of the equations (\ref{w83}) need not exist. 
In order to study its existence, let us consider the condition (iii) in 
Proposition \ref{w39}. 

\begin{theo} \label{w120} \mar{w120}
The graded Lagrangian $L_e$ (\ref{w8}) can be extended to a solution
(\ref{w61}) of the master equation only if the graded derivation
$u_e$ (\ref{w108}) is extended to a graded derivation nilpotent on the
shell.
\end{theo}

\begin{proof}
Given a graded Lagrangian (\ref{w61}), 
the corresponding graded
derivations (\ref{w37}) read
\mar{w102,3}\ben
&& \up=  \frac{\op\dl^\lto (\cL_1+\cL')}{\dl \ol
s_A}\frac{\dr}{\dr s^A} + \op\sum_{0\leq k\leq N}
\frac{\op\dl^\lto (\cL_1+\cL')}{\dl \ol
c_{r_k}}\frac{\dr}{\dr c^{r_k}}, \label{w102} \\
&& \ol\up= \frac{\rdr }{\dr \ol
s_A}\frac{\dl(\cL+\cL_1+\cL')}{\dl s^A} +
\op\sum_{0\leq k\leq N}
\frac{\rdr }{\dr \ol
c_{r_k}}\frac{\dl (\cL_1+\cL')}{\dl c^{r_k}}. \label{w103}
\een
Then the condition (iii) can be written in the form (\ref{0688}) as
\mar{w131}\beq
(\up +\ol\up)(\up)=0, \qquad (\up +\ol\up)(\ol\up)=0. \label{w131}
\eeq
It falls into a set of equalities with respect to the polynomial 
degree in antifields. Let us put
\be
\up=\up^0+\up^1 +\up', \qquad \ol\up=\ol\up^0 +\ol\up',
\ee 
where $\up^0$ and $\up^1$ are the parts of $\up$ (\ref{w102}) 
of zero and first polynomial degree in antifields, respectively,
and $\ol\up^0$ is that of $\ol\up$ (\ref{w103}) 
independent of antifields. It is readily observed that $\ol\up^0=\ol\dl$
is the Koszul--Tate differential.
Let us consider the part of the equalities (\ref{w131}) which is
independent of antifields. It reads
\mar{w104}\beq
\up^0(\up^0) +\ol\up^0(\up^1)= \up^0(\up^0) +\ol\dl(\up^1)=0,
\label{w104}
\eeq
i.e.,
the graded derivation $\up^0$ vanishes on the shell. It is 
readily observed that the part of $\up^0$ linear in ghosts is exactly
the total gauge operator $u_e$ (\ref{w108}), i.e., $\up^0$ provides a
nilpotent extension $u_e$ on the shell.
\end{proof}

In other words, the Lagrangian  $L_e$ (\ref{w8}) is extended 
to a solution of the master equation only if the gauge and higher-stage
gauge supersymmetries of an original Lagrangian $L$ form an algebra
on the shell.

In order to formulate the sufficient condition, let us assume 
that the gauge and higher-stage
gauge supersymmetries of an original Lagrangian $L$ form an algebra,
and this algebra is given from the beginning, i.e., we have 
a nilpotent extension $\up^0$ of 
the total gauge operator $u_e$ (\ref{w108}).

\begin{theo} \label{w130} \mar{w130}
Let the graded derivation $u_e$ (\ref{w108}) admit a nilpotent
extension $\up^0$ (\ref{w109}) of zero antifield number. Then the graded
Lagrangian 
\mar{w133}\beq
\gL\om=L_e + \op\sum_{1\leq k\leq N}\xi^{r_{k-1}}\ol c_{r_{k-1}}\om
\label{w133}
\eeq
satisfies the master equation.
\end{theo}

\begin{proof}
If the graded derivation $\up^0$ is nilpotent, then $\ol\dl(\up^1)=0$ by
virtue of the equation (\ref{w104}). It follows that the part $L^2_1$ of
the Lagrangian
$L_e$ quadratic in antifields obeys the relations
$\ol\dl(\op\dl^\lto{}^{r_k}(\cL^2_1))=0$ for all indices $r_k$. This part
consists of the terms
$h_{r_k}^{(r_{k-2},\Si)(A,\Xi)}\ol c_{\Si r_{k-2}}\ol s_{\Xi A}$
(\ref{v92'}), which consequently are
$\ol\dl$-closed. Then the summand $G_{r_k}$ of each cocycle
$\Delta_{r_k}$ (\ref{v92'}) is $\dl_{k-1}$-closed in accordance with the
relation (\ref{v162}). It follows
that its summand $h_{r_k}$ is also $\dl_{k-1}$-closed and, consequently,
$\dl_{k-2}$-closed. Hence it is $\dl_{k-1}$-exact by virtue the
homology regularity condition. Therefore, $\Delta_{r_k}$ is
reduced only to the  summand $G_{r_k}$ linear in antifields. It follows
that the Lagrangian $L_1$ (\ref{w8}) is linear in antifields. In this
case, we have
\be
u^A=\op\dl^\lto{}^A(\cL_e), \qquad u^{r_k}=\op\dl^\lto{}^{r_k}(\cL_e)
\ee
for all indices $A$ and $r_k$ and, consequently, 
\be
\up^A=\op\dl^\lto{}^A(\gL), \qquad \up^{r_k}=\op\dl^\lto{}^{r_k}(\gL),
\ee
i.e., $\up^0$ is the graded derivation $\up$ (\ref{w37}) defined by
the Lagrangian (\ref{w133}). Then the nilpotency condition
$\up^0(\up^0)=0$ takes the form
\be
\up^0(\op\dl^\lto{}^A(\gL))=0, \qquad \up^0(\op\dl^\lto{}^{r_k}(\gL))=0.
\ee
Hence, we obtain
\be
\up^0(\gL)=\up^0(\op\dl^\lto{}^A(\gL)\ol s_A + \op\dl^\lto{}^{r_k}(\gL)
\ol c_{r_k})=0,
\ee
i.e., $\up^0$ is a variational supersymmetry of the Lagrangian
(\ref{w133}).  Thus, it satisfies the master equation in accordance
with Proposition \ref{w39}
\end{proof}

Since the summand $L_1$ of the Lagrangian (\ref{w133}) is linear in
antifields, the Lagrangian (\ref{w133}) up to a $d_H$-exact term can be
written in the form
\mar{w300}\beq
\gL\om= L +u^A\ol s_A +
\op\sum_{1\leq k\leq N}(u^{r_{k-1}} +\xi^{r_{k-1}})\ol c_{r_{k-1}}\om,
\label{w300} 
\eeq
which is also a solution of the master equation.

\section{Example}

We address the
topological BF theory of two exterior forms $A$ and $B$ of form degree
$|A|+|B|=\di X-1$ on a smooth manifold $X$ \cite{birm}. It is a
reducible degenerate Lagrangian theory \cite{jpa05}. 
Since the verification of the homology regularity condition in a general
case is rather complicated, we here restrict our consideration to 
the simplest example of the topological BF theory when $A$ is a
function \cite{jmp05a}. 

Let us consider the fiber bundle
\be
Y=\Bbb R\op\times_X \op\w^{n-1} T^*X,
\ee
coordinated by $(x^\la, A, B_{\m_1\ldots \m_{n-1}})$ and provided with 
the canonical $(n-1)$-form
\be
B=\frac{1}{(n-1)!}B_{\m_1\ldots \m_{n-1}}dx^{\m_1}\w\cdots\w
dx^{\m_{n-1}}.
\ee
The Lagrangian and the Euler--Lagrange operator of the topological BF
theory in question read
\mar{v182,3}\ben
&& L_{\rm BF}=\frac1n Ad_HB, \label{v182}\\
&&\dl L= dA\w \cE\om + dB_{\m_1\ldots \m_{n-1}}\w \cE^{\m_1\ldots
\m_{n-1}}\om, \nonumber\\
&& \cE=\e^{\m\m_1\ldots \m_{n-1}} d_\m B_{\m_1\ldots \m_{n-1}},
\qquad \cE^{\m_1\ldots \m_{n-1}} = - \e^{\m\m_1\ldots
\m_{n-1}}d_\m A, \label{v183}
\een
where $\e$ is the Levi--Civita symbol.

Let us extend the BGDA $\cO^*_\infty Y$ to the BGDA
$\cP^*_\infty[\ol Y^*;Y]$ where
\be
VY=Y\op\times_X Y, \qquad \ol Y^*= (\Bbb R\op\times_X
\op\w^{n-1}TX)\op\ot_X \op\w^n T^*X.
\ee
This BGDA  possesses the local basis $\{ A, B_{\m_1\ldots
\m_{n-1}}, \ol s, \ol s^{\m_1\ldots \m_{n-1}}\}$, where $\ol s,
\ol s^{\m_1\ldots \m_{n-1}}$ are odd antifields of antifield number 1.
With the nilpotent Koszul--Tate differential
\be
\ol\dl=\frac{\rdr}{\dr \ol s}\cE + \frac{\rdr}{\dr \ol
s^{\m_1\ldots \m_{n-1}}} \cE^{\m_1\ldots \m_{n-1}},
\ee
we have the complex (\ref{v042}),
\be
0\lto \im\ol\dl \llr^{\ol\dl} \cP^{0,n}_\infty[\ol Y^*;Y]_1
\llr^{\ol\dl} \cP^{0,n}_\infty[\ol Y^*;Y]_2.
\ee
A generic one-chain reads
\be
\Phi= \op\sum_{0\leq |\La|}(\Phi^\La\ol s_\La +
\Phi^\La_{\m_1\ldots \m_{n-1}} \ol s^{\m_1\ldots
\m_{n-1}}_\La)\om,
\ee
and the cycle condition $\ol\dl\Phi=0$ takes the form
\mar{v189}\beq
\Phi^\La\cE_\La + \Phi^\La_{\m_1\ldots \m_{n-1}} \cE^{\m_1\ldots
\m_{n-1}}_\La=0. \label{v189}
\eeq
If $\Phi^\La$ and $\Phi^\La_{\m_1\ldots\m_{n-1}}$ are independent
of the variational derivatives (\ref{v183}) (i.e., $\Phi$ is a
nontrivial cycle), the equality (\ref{v189}) is split into the
following: 
\be
\Phi^\La\cE_\La=0, \qquad
 \Phi^\La_{\m_1\ldots \m_{n-1}} \cE^{\m_1\ldots
\m_{n-1}}_\La=0.
\ee
The first equality holds iff $\Phi^\La=0$, i.e., there is
no Noether identity involving $\cE$. The second one is
satisfied iff
\be
\Phi^{\la_1\ldots \la_k}_{\m_1\ldots\m_{n-1}}\e^{\m\m_1\ldots
\m_{n-1}}=- \Phi^{\m\la_2\ldots
\la_k}_{\m_1\ldots\m_{n-1}}\e^{\la_1\m_1\ldots \m_{n-1}}.
\ee
It follows that $\Phi$ factorizes as
\be
\Phi= \op\sum_{0\leq |\Xi|} G_{\nu_2\ldots\nu_{n-1}}^\Xi
d_\Xi\Delta^{\nu_2\ldots\nu_{n-1}}\om
\ee
via local graded densities
\mar{v190}\beq
\Delta^{\nu_2\ldots\nu_{n-1}}=\Delta^{\nu_2\ldots\nu_{n-1},
\la}_{\al_1\ldots\al_{n-1}}\ol
s^{\al_1\ldots\al_{n-1}}_\la=\dl^\la_{\al_1}\dl^{\nu_2}_{\al_2}\cdots
\dl^{\nu_{n-1}}_{\al_{n-1}} \ol s^{\al_1\ldots\al_{n-1}}_\la=
d_{\nu_1}\ol s^{\nu_1\nu_2\ldots\nu_{n-1}}, \label{v190}
\eeq
which provide the complete Noether identities
\mar{v191}\beq
d_{\nu_1}\cE^{\nu_1\nu_2\ldots\nu_{n-1}}=0. \label{v191}
\eeq

The local graded densities (\ref{v190}) form the basis for a
projective $C^\infty(X)$-module of finite rank which is isomorphic
to the module of sections of the vector bundle
\be
\ol V^*=\op\w^{n-2} TX\op\ot_X \op\w^n T^*X, \qquad V= \op\w^{n-2}
T^*X.
\ee
Therefore, let us extend the BGDA $\cP^*_\infty[\ol Y^*;Y]$ to the
BGDA $\cP^*_\infty\{0\}= \cP^*_\infty[\ol Y^*;Y;V]$ possessing the
local basis
$\{A, B_{\m_1\ldots \m_{n-1}}, \ol s, \ol s^{\m_1\ldots \m_{n-1}},
\ol c^{\m_2\ldots \m_{n-1}}\}$,
where $\ol
c^{\m_2\ldots \m_{n-1}}$ are even antifields of antifield number 2. We
have the nilpotent graded derivation
\be
\dl_0= \ol\dl + \frac{\rdr}{\dr \ol c^{\m_2\ldots \m_{n-1}}}
\Delta^{\m_2\ldots \m_{n-1}}
\ee
of $\cP^*_\infty\{0\}$. Its nilpotency
is equivalent to the Noether identities (\ref{v191}). Then we obtain the
one-exact complex
\be
0\lto \im\ol\dl \llr^{\ol\dl} \cP^{0,n}_\infty[\ol Y^*;Y]_1
\llr^{\dl_0} \cP^{0,n}_\infty\{0\}_2 \llr^{\dl_0}
\cP^{0,n}_\infty\{0\}_3.
\ee

Iterating the arguments, we come to the 
$(N+1)$-exact complex (\ref{v94}) for $N\leq n-3$ as follows.
Let us consider the vector bundles
\be
V_k=\op\w^{n-k-2} T^*X, \qquad k=1,\ldots, N,
\ee
and the corresponding BGDA
$\cP^*_\infty\{N\}= \cP^*_\infty[...V_3V_1\ol Y^*;Y;VV_2V_4...]$,
possessing the local basis
\be
&&\{A, B_{\m_1\ldots \m_{n-1}}, \ol s, \ol s^{\m_1\ldots
\m_{n-1}}, \ol c^{\m_2\ldots \m_{n-1}},\ldots,\ol c^{\m_{N+2}\ldots \m_{n-1}}\},\\
&& [\ol c^{\m_{k+2}\ldots \m_{n-1}}]=(k+1){\rm mod}\,2, \qquad
{\rm Ant}[\ol c^{\m_{k+2}\ldots \m_{n-1}}]=k+3.
\ee
It is provided with the nilpotent graded derivation
\mar{va202}\beq
\dl_N=\dl_0 + \op\sum_{1\leq k\leq N}\frac{\rdr}{\dr \ol
c^{\m_{k+2}\ldots \m_{n-1}}}
 \Delta^{\m_{k+2}\ldots \m_{n-1}}, \qquad
\Delta^{\m_{k+2}\ldots
\m_{n-1}}=d_{\m_{k+1}}\ol c^{\m_{k+1}\m_{k+2}\ldots \m_{n-1}},
\label{va202}
\eeq
of antifield number -1. Its nilpotency results from the Noether
identities (\ref{v191}) and the equalities
\mar{v212}\beq
d_{\m_{k+2}}\Delta^{\m_{k+2}\ldots \m_{n-1}}=0, \qquad
k=0,\ldots,N, \label{v212}
\eeq
which are $k$-stage Noether identities \cite{jpa05}. Then the
manifested $(N+1)$-exact complex reads
\mar{v203}\ben
&&0\lto \im \ol\dl \llr^{\ol\dl} \cP^{0,n}_\infty[\ol
Y^*;Y]_1\llr^{\dl_0} \cP^{0,n}_\infty\{0\}_2\llr^{\dl_1}
\cP^{0,n}_\infty\{1\}_3\cdots
\label{v203}\\
&& \qquad
 \llr^{\dl_{N-1}} \cP^{0,n}_\infty\{N-1\}_{N+1}
\llr^{\dl_N} \cP^{0,n}_\infty\{N\}_{N+2}\llr^{\dl_N}
\cP^{0,n}_\infty\{N\}_{N+3}. \nonumber
\een
It obeys the following $(N+2)$-homology regularity condition (see
Appendix B for the proof).

\begin{lem} \label{v220} \mar{v220}
Any $(N+2)$-cycle $\Phi\in \cP^{0,n}_\infty\{N-1\}_{N+2}$ up to a
$\dl_{N-1}$-boundary takes the form
\mar{v218}\ben
&& \Phi=\op\sum_{k_1+\cdots +k_i+3i=N+2}\sum_{0\leq|\La_1|,\ldots,
|\La_i|}G^{\La_1\cdots \La_i}_{\m^1_{k_1+2}\ldots
\m^1_{n-1};\ldots; \m^i_{k_i+2}\ldots \m^i_{n-1}} \label{v218} \\
&&\qquad d_{\La_1} \Delta^{\m^1_{k_1+2}\ldots \m^1_{n-1}}\cdots
d_{\La_i} \Delta^{\m^i_{k_i+2}\ldots \m^i_{n-1}}\om, \qquad
k=-1,0,1,\ldots, N, \nonumber
\een
where $k=-1$ stands for
$\ol c^{\m_1\ldots\m_{n-1}}=\ol s^{\m_1\ldots\m_{n-1}}$ and 
$\Delta^{\m_1\ldots\m_{n-1}}=\cE^{\m_1\ldots\m_{n-1}}$.
It follows that $\Phi$ is a $\dl_N$-boundary.
\end{lem}

Following the proof of Lemma \ref{v220}, one can also show that any
$(N+2)$-cycle $\Phi\in \cP^{0,n}_\infty\{N\}_{N+2}$ up to a
boundary takes the form
\be
\Phi=\op\sum_{0\leq|\La|}G^\La_{\m_{N+2}\ldots \m_{n-1}} d_\La
\Delta^{\m_{N+2}\ldots \m_{n-1}}\om, 
\ee
i.e., the homology $H_{N+2}(\dl_N)$ of the complex (\ref{v203}) is
finitely generated by the cycles $\Delta^{\m_{N+2}\ldots
\m_{n-1}}$. Thus, the complex (\ref{v203}) admits the
$(N+2)$-exact extension (\ref{v171}).

The iteration procedure is prolonged till $N=n-3$. We have the
BGDA $\cP^*\{n-2\}$, where $V_{n-2}=X\times\Bbb R$.
It possesses the local basis
\be
\{A, B_{\m_1\ldots \m_{n-1}}, \ol s, \ol s^{\m_1\ldots \m_{n-1}},
\ol c^{\m_2\ldots \m_{n-1}},\ldots,\ol c^{\m_{n-1}}, \ol c\},
\ee
where $[\ol c]=(n-1){\rm mod}\,2$ and Ant$[\ol c]=n+1$. 
The corresponding Koszul--Tate complex reads
\be
&&0\lto \im \ol\dl \llr^{\ol\dl} \cP^{0,n}_\infty[\ol
Y^*;Y]_1\llr^{\dl_0} \cP^{0,n}_\infty\{0\}_2\llr^{\dl_1}
\cP^{0,n}_\infty\{1\}_3\cdots\\
&& \qquad
 \llr^{\dl_{n-3}} \cP^{0,n}_\infty\{n-3\}_{n-1}
\llr^{\dl_{n-2}} \cP^{0,n}_\infty\{n-2\}_n\llr^{\dl_{n-2}}
\cP^{0,n}_\infty\{n-2\}_{n+1}. \\
&&\dl_{n-2}=\dl_0 + \op\sum_{1\leq k\leq n-3}\frac{\rdr}{\dr \ol
c^{\m_{k+2}\ldots \m_{n-1}}} \Delta^{\m_{k+2}\ldots \m_{n-1}} +
\frac{\rdr}{\dr \ol c}\Delta, \qquad \Delta=d_{\m_{n-1}}\ol
c^{\m_{n-1}}.
\ee

Let us enlarge the BGDA $\cP^*\{n-2\}$ to the BGDA $P^*\{n-2\}$ (\ref{w6})
possessing the local basis
\be
\{A, B_{\m_1\ldots \m_{n-1}}, c_{\m_2\ldots \m_{n-1}},\ldots,
c_{\m_{n-1}},  c,
 \ol s, \ol s^{\m_1\ldots \m_{n-1}},
\ol c^{\m_2\ldots \m_{n-1}},\ldots,\ol c^{\m_{n-1}}, \ol c\},
\ee
where $c_{\m_2\ldots \m_{n-1}},\ldots,
c_{\m_{n-1}},  c$ are the corresponding ghosts. Let us extend the
Lagrangian $L_{\rm BF}$ (\ref{v182}) to the even graded density
\mar{w302}\ben
&& L_e= L_{\rm BF} + L_1=L_{\rm BF} +  [\op\sum_{0\leq k\leq n-3}
c_{\m_{k+2}\ldots \m_{n-1}} \Delta^{\m_{k+2}\ldots \m_{n-1}} +
 c\Delta] \om= \label{w302}\\
&& \quad L_{\rm BF} + [c_{\m_2\ldots \m_{n-1}} d_{\m_1} \ol
s^{\m_1\m_2\ldots
\m_{n-1}} + \op\sum_{1\leq k\leq n-3}
c_{\m_{k+2}\ldots \m_{n-1}} d_{\m_{k+1}} \ol c^{\m_{k+1}\m_{k+2}\ldots
\m_{n-1}} +c d_{\m_{n-1}}\ol
c^{\m_{n-1}}]\om. \nonumber
\een
Since the graded density $L_1$ is independent on $A$ and $B_{\m_1\ldots
\m_{n-1}}$, the relation (\ref{w72}) holds and, therefore, $L_e$
(\ref{w302}) is a solution of the master equation.

\section{Appendix A}

We start the proof of Theorem \ref{v11} with the following algebraic
Poincar\'e lemma.

\begin{lem} \label{v38'} \mar{v38'}
If $Y=\Bbb R^{n+m}\to \Bbb R^n$, the complex (\ref{g111}) at all
the terms, except $\Bbb R$, is exact, while the complex
(\ref{g112}) is exact.
\end{lem}

\begin{proof}
This is the case of an affine bundle $Y$, and the above mentioned
exactness has been proved when the ring $\cO^0_\infty Y$ is
restricted to the subring $\cP^0_\infty Y$ of polynomial functions
(see \cite{cmp04}, Lemmas 4.2 -- 4.3). The proof of these lemmas
is straightforwardly extended to $\cO^0_\infty Y$ if the homotopy
operator (4.5) in \cite{cmp04}, Lemma 4.2 is replaced with that
(4.8) in \cite{cmp04}, Remark 4.1.
\end{proof}

The proof of Theorem \ref{v11} follows that of \cite{cmp04}, Theorem 2.1.
We first prove Theorem \ref{v11} for the above mentioned BGDA
$\G(\gQ^*_\infty[F;Y])$. Similarly to $\cS^*_\infty[F;Y]$, the
sheaf $\gQ^*_\infty[F;Y]$ and the BGDA $\G(\gQ^*_\infty[F;Y])$ are
split into the variational bicomplexes, and we consider their
subcomplexes
\mar{v35-8}\ben
&& 0\ar \Bbb R\ar \gQ^0_\infty[F;Y]\ar^{d_H}\gQ^{0,1}_\infty[F;Y]
\cdots \ar^{d_H} \gQ^{0,n}_\infty[F;Y]\ar^\dl {\got E}_1,
\label{v35}\\
&& 0\to \gQ^{1,0}_\infty[F;Y]\ar^{d_H} \gQ^{1,1}_\infty[F;Y]\cdots
\ar^{d_H}\gQ^{1,n}_\infty[F;Y]\ar^\vr {\got E}_1\to 0, \label{v36}\\
&& 0\ar \Bbb R\ar
\G(\gQ^0_\infty[F;Y])\ar^{d_H}\G(\gQ^{0,1}_\infty[F;Y]) \cdots
\ar^{d_H} \G(\gQ^{0,n}_\infty[F;Y])\ar^\dl \G({\got E}_1),
\label{v37} \\
&&  0\to \G(\gQ^{1,0}_\infty[F;Y])\ar^{d_H}
\G(\gQ^{1,1}_\infty[F;Y])\cdots
\ar^{d_H}\G(\gQ^{1,n}_\infty[F;Y])\ar^\vr \G({\got E}_1)\to 0,
\label{v38}
\een
where ${\got E}_1 =\vr(\gQ^{1,n}_\infty[F;Y])$. By virtue of Lemma
\ref{v38'}, the complexes (\ref{v35}) -- (\ref{v36}) at all the
terms, except $\Bbb R$, are exact. The terms
$\gQ^{*,*}_\infty[F;Y]$ of the complexes (\ref{v35}) --
(\ref{v36}) are sheaves of $\G(\gQ^0_\infty)$-modules. Since
$J^\infty Y$ admits a partition of unity just by elements of
$\G(\gQ^0_\infty)$, these sheaves are fine and, consequently,
acyclic. By virtue of the abstract de Rham theorem (see
\cite{cmp04}, Theorem 8.4, generalizing \cite{hir}, Theorem
2.12.1), cohomology of the complex (\ref{v37}) equals the
cohomology of $J^\infty Y$ with coefficients in the constant sheaf
$\Bbb R$ and, consequently, the de Rham cohomology of $Y$, which
is the strong deformation retract of $J^\infty Y$. Similarly, the
complex (\ref{v38}) is proved to be exact. It remains to prove
that cohomology of the complexes (\ref{g111}) -- (\ref{g112})
equals that of the complexes (\ref{v37}) -- (\ref{v38}). The proof
of this fact straightforwardly follows the proof of \cite{cmp04},
Theorem 2.1, and it is a slight modification of the proof of
\cite{cmp04}, Theorem 4.1, where graded exterior forms on the
infinite order jet manifold $J^\infty Y$ of an affine bundle are
treated as those on $X$.

\section{Appendix B}

In order to prove Lemma \ref{v220},
let us choose some basis element $\ol c^{\m_{k+2}\ldots \m_{n-1}}$
and denote it simply by $\ol c$. Let $\Phi$ contain a summand
$\f_1 \ol c$, linear in $\ol c$. Then the cycle condition reads
\be
\dl_{N-1}\Phi=\dl_{N-1}(\Phi-\f_1 \ol c) + (-1)^{[\ol
c]}\dl_{N-1}(\f_1)\ol c + \f \Delta=0, \qquad \Delta=\dl_{N-1}\ol
c.
\ee
It follows that $\Phi$ contains a summand $\psi\Delta$ such that
\be
(-1)^{[\ol c]+1}\dl_{N-1}(\psi)\Delta +\f\Delta=0.
\ee
This equality implies the relation
\mar{v213}\beq
\f_1=(-1)^{[\ol c]+1}\dl_{N-1}(\psi) \label{v213}
\eeq
because the reduction conditions (\ref{v212}) involve total
derivatives of $\Delta$, but not $\Delta$. Hence,
\be
\Phi=\Phi' +\dl_{N-1}(\psi \ol c),
\ee
where $\Phi'$ contains no term linear in $\ol c$. Furthermore, let
$\ol c$ be even and $\Phi$ has a summand $\sum \f_r \ol c^r$
polynomial in $\ol c$. Then the cycle condition leads to the
equalities
\be
\f_r\Delta=-\dl_{N-1}\f_{r-1}, \qquad r\geq 2.
\ee
Since $\f_1$ (\ref{v213}) is $\dl_{N-1}$-exact, then $\f_2=0$ and,
consequently, $\f_{r>2}=0$. Thus, a cycle $\Phi$ up to a
$\dl_{N-1}$-boundary contains no term polynomial in $\ol c$. It
reads
\mar{v217}\beq
\Phi=\op\sum_{k_1+\cdots +k_i+3i=N+2}\sum_{0<|\La_1|,\ldots,
|\La_i|}G^{\La_1\cdots \La_i}_{\m^1_{k_1+2}\ldots
\m^1_{n-1};\ldots; \m^i_{k_i+2}\ldots \m^i_{n-1}} \ol
c^{\m^1_{k_1+2}\ldots \m^1_{n-1}}_{\La_1}\cdots \ol
c_{\La_i}^{\m^i_{k_i+2}\ldots \m^i_{n-1}}\om. \label{v217}
\eeq
However, the terms polynomial in $\ol c$ may appear under general
covariant transformations
\be
\ol c'^{\nu_{k+2}\ldots \nu_{n-1}}=\det(\frac{\dr x^\al}{\dr
x'^\bt}) \frac{\dr x'^{\nu_{k+2}}}{\dr x^{\m_{k+2}}}\cdots
\frac{\dr x'^{\nu_{n-1}}}{\dr x^{\m_{n-1}}}\ol c^{\m_{k+2}\ldots
\m_{n-1}}
\ee
of a chain $\Phi$ (\ref{v217}). In particular, $\Phi$ contains the
summand
\be
\op\sum_{k_1+\cdots +k_i+3i=N+2}F_{\nu^1_{k_1+2}\ldots
\nu^1_{n-1};\ldots; \nu^i_{k_i+2}\ldots \nu^i_{n-1}} \ol
c'^{\nu^1_{k_1+2}\ldots \nu^1_{n-1}}\cdots \ol
c'^{\nu^i_{k_i+2}\ldots \nu^i_{n-1}},
\ee
which must vanish if $\Phi$ is a cycle. This takes place only if
$\Phi$ factorizes through the graded densities
$\Delta^{\m_{k+2}\ldots \m_{n-1}}$ (\ref{va202}) in accordance
with the expression (\ref{v218}).

\end{document}